\documentclass[twocolumn,superscriptaddress,preprintnumbers,aps]{revtex4}

\usepackage{amsmath}
\usepackage{graphicx}

\usepackage{hyperref}
\hypersetup{
    colorlinks=true,       
    linkcolor=blue,        
    citecolor=blue,        
    filecolor=blue,        
    urlcolor=blue,         
    runcolor=blue
}

    \def\bra#1{\langle {#1} |}
    \def\ket#1{| {#1} \rangle}
\def\E{\mathbf{E}}
\def\r{\mathbf{r}}
\def\P{\mathbf{P}}
\def\D{\mathbf{D}}

\usepackage{xcolor}
\newcommand{\notablue}[1]{{\color{black}\textrm{#1}}}

\newcommand{\notaviolet}[1]{{\color{black}\textrm{#1}}}

\begin{document}

\title{Disordered ensembles of strongly coupled single-molecule plasmonic picocavities\\ as nonlinear optical metamaterials}

\author{Felipe Herrera}
\email{felipe.herrera.u@usach.cl}
\affiliation{Department of Physics, Universidad de Santiago de Chile, Av. Ecuador 3493, Santiago, Chile.}
\affiliation{ANID-Millennium Institute for Research in Optics, Chile.}

\author{Marina Litinskaya}
\email{litinskaya@gmail.com}
\affiliation{Department of Physics \& Astronomy, University of British Columbia, Vancouver, Canada, V6T 1Z1}

\date{\today}

\begin{abstract}
\notablue{We propose to use molecular picocavity ensembles as macroscopic coherent nonlinear optical devices enabled by nanoscale strong coupling. For a generic picocavity model that includes molecular and photonic disorder, we derive theoretical performance bounds for coherent cross-phase modulation signals using weak classical fields of different frequencies. We show that strong coupling of the picocavity {\it vacua}  with a specific vibronic sideband in the molecular emission spectrum results in a significant variation of the effective refractive index of the metamaterial relative to a molecule-free scenario, due to a vacuum-induced Autler-Townes effect. For a realistic molecular disorder model, we demonstrate that cross-phase modulation of optical fields as weak as 10 kW/cm$^2$ is feasible using dilute ensembles of molecular picocavities at room temperature, provided that the confined vacuum is not resonantly driven by the external probe field}. Our work paves the way for the development of plasmonic metamaterials that exploit strong coupling for optical state preparation and quantum control.
\end{abstract}

\maketitle

\section{Introduction}
Strong light-matter coupling with single molecules in plasmonic picocavities has emerged as a  resource for room-temperature quantum control with nanoscale optical fields. Organic chromophores in plasmonic picocavities \cite{Benz2016,Carnegie2018,Chikkaraddy:2016aa,Chikkaraddy2018} are promising  platforms for studying cavity quantum electrodynamics (QED) at room temperature \cite{Ebbesen2016,Herrera2020perspective}. Recent experiments \cite{Zhang2021} and rigorous theoretical modeling \cite{Delga2014,Neuman2018,Neuman2020,Feist2021} have emphasized the quantum optical origin of commonly used plasmon-enhanced molecular spectroscopy techniques \cite{Schmid2013}, offering new perspectives on conventional architectures that can stimulate the study of novel schemes for optical quantum control at the nanoscale \cite{Tame2013}.

Conventional molecular-cavity QED platforms based on planar optical microcavities exploit the interaction of the electromagnetic vacuum with an ensemble of vibronic coherences to reach the strong and ultrastrong coupling regimes \cite{Agranovich2005,Tischler2007,kena-cohen2008,Herrera2017-PRA,Herrera2017-PRL,Mazzeo2014,Gambino2015,Schartz2011,Kena-Cohen2013,Hobson2002}. The collective character of the interaction can enhance electric and charge transport processes \cite{Feist2015,Schachenmayer2015,yuen2016,Pupillo2017}, mediated by the cavity-induced delocalization of the molecular degrees of freedom involved in the transport process. Collective coupling can also lead to modifications of the chemical reactivity \cite{Hutchison2012,Herrera2016} and optical response \cite{Herrera2014,Barachati2018,Daskalakis2017,Lerario2017} of organic materials. Collective strong coupling in microcavity occurs through a mechanism analogous to dipole synchronization \cite{Zhu:2015}. However, the local field that each individual molecule experiences in a microcavity is relatively small.

In contrast, in plasmonics, the extreme sub-wavelength field confinement achievable with current technology  \cite{Baranov2017,Muller2018} allows for light-matter interaction energies to overcome  local thermal fluctuations, at the level of individual molecules. This can enable the implementation of local control protocols that exploit strong vacuum fields for studying  optomechanical physics \cite{Zhang2021} and tailored photochemistry \cite{Felicetti2020,Fregoni2018}. Nanoparticle fabrication techniques can produce a large number of ``molecular picocavities" \cite{Carnegie2018}. The picocavity distribution can be strongly inhomogeneous, and must be sampled locally using tip-based nanoprobes  \cite{Behr2008,May2020,Metzger2019}, to extract spatially-resolved information about the light-matter coupling dynamics in the system. Although chemical methods are available to increase the homogeneity and reproducibility of the picocavity fabrication \cite{Chikkaraddy2018}, the need to develop efficient local sampling method may be a challenge for the scalability and integrability of molecular picocavities in next-generation nanophotonic devices.

Instead of focusing on the local aspects of light-matter interaction in disordered picocavity ensembles, in Ref. \cite{Litinskaya2019} we explored a macroscopic approach in which cavity strong coupling was used for inducing nonlinear optical signals in the response of the ensemble. We assumed a scheme involving  {\it cis}-{\it trans} molecular isomers (photoswitches) that are embedded in high-quality optical microcavities with photon lifetimes of several picoseconds. We exploited the unique spectral and coherence properties of molecular photoswitches to find suitable conditions for inducing a cavity-assisted transparency window in the absorption spectrum and implementing cross-phase modulation between external laser fields. The phase nonlinearity was shown to be robust with respect to static disorder in the molecular dipole orientation and molecular transition frequencies, given a set of restrictions on the allowed vibrational and photonic dephasing times.

In this work, we significantly generalize the analysis in Ref. \cite{Litinskaya2019} that could facilitate experimental implementations. We achieve this by reducing the number of physical assumptions imposed over the relevant molecular and photonic degrees of freedom, in particular, the type and properties of the cavity resonator structures needed for field confinement, and the class of organic molecules that couple to the vacuum cavity field. We now consider a broad class of organic chromophores that exhibit significant electron-vibration coupling in the lowest electronic singlet transition $S_0\leftrightarrow S_1$ for an intramolecular vinyl stretching mode of frequency $\omega_{\rm v}\approx 0.2$ eV~\cite{Spano2011}. \notablue{We assume that static fluctuations of the molecular transition frequencies are the leading source of inhomogeneity in the ensemble, in an effort to understand the fundamental limits of coherent optical signals in a scenario where the energy disorder is dominant. Other types of inhomogeneity such as random dipole orientations have a smaller effect on the optical response of coupled light-matter systems \cite{Litinskaya2019}. }

\notablue{On the photonic side, we adopt three simplifying assumptions about the properties of picocavities: (\emph{i}) We consider a dilute ensemble of picocavity structures that are much smaller than the optical probe wavelength, and have negligible inter-particle interaction. This allows us to treat the ensemble as an effective medium which to  lowest order is dominated by the single-particle response; (\emph{ii}) The near-field spectrum of an individual empty picocavity is treated as a single Lorentzian feature that is red-detuned from the external probe frequency and has a bandwidth (FHWM) not greater than the molecular vibration frequency. This ensures that the direct laser excitation of the picocavity gap resonance is suppressed for weak probe field intensities; (\emph{iii}) Strong coupling is achieved on average within each single-molecule picocavity. 
Our results rest on these three conditions being simultaneously satisfied, which may be challenging to realize with currently available plasmonic picocavities  \cite{Benz2016,Carnegie2018,Chikkaraddy:2016aa,Chikkaraddy2018}. } 

\notablue{Building on these assumptions, we provide a  proof-of-principle demonstration} that single-molecule strong coupling with the picocavity vacuum induces phase modulation of a probe laser field that coherently drives a disordered ensemble with varying light-matter coupling strengths, inhomogeneous broadening of molecular transition frequencies, and sub-picosecond decoherence of molecular and photonic degrees of freedom. We also show that the presence of a second (signal) laser induces a controllable vacuum-enhanced cross-phase modulation, at intensities as low as a few kW/cm$^2$. Such intensities are orders of magnitude lower than the light sources used in conventional nonlinear optics \cite{Bhawalkar1996}. The ensemble nonlinearity can be optically gated and used as an optical switch, exploiting coherence at macroscopic scales despite the strong structural disorder.

In the rest of this article, we introduce the macroscopic approach that allows us to define the  effective optical susceptibility of an ensemble of molecular picocavities (Sec. \ref{sec:effective index}) and a nonlinear phase modulation signal (Sec. \ref{sec:phase shifts}). Then we discuss the quantum electrodynamics model that describes the light-matter dynamics in individual picocavities (Sec. \ref{sec:level scheme}), and finally study the dependence of the nonlinear phase modulation observables on the system parameters (Sec. \ref{sec:homogeneous AT}-\ref{sec:inhomogeneous AT}). We summarize our main results in Sec. \ref{sec:conclusions}.

\section{Effective Medium Approach}
\label{sec:picocavities}

\begin{figure}[t]
\includegraphics[width=0.5\textwidth]{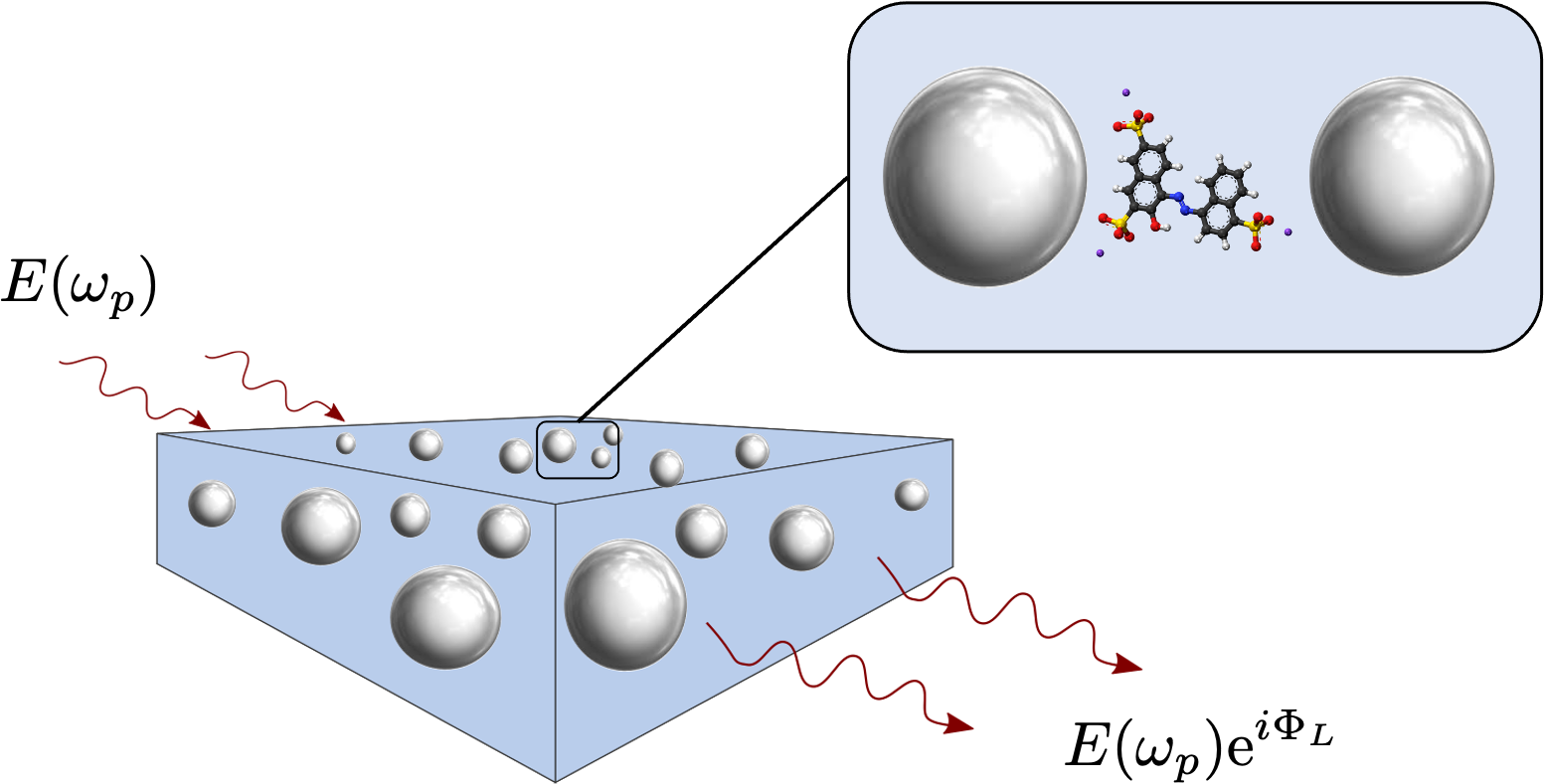}
\caption{{\bf Single-molecule picocavity metasurface}. Large ensemble of metal nanoparticle inclusions on a thin dielectric layer subject to external in-plane driving by a probe field at frequency $\omega_p$. Each plasmonic picocavity in the ensemble contains an individual organic molecule in the gap region where the electromagnetic field is strongly confined. Optical interactions with the molecules results in a phase shift $\Phi_L$ of the incoming wave, measured relative to the molecule-free layer. }
\label{fig:metasurface}
\end{figure}

\begin{figure*}[t]
\includegraphics[width=0.95\textwidth]{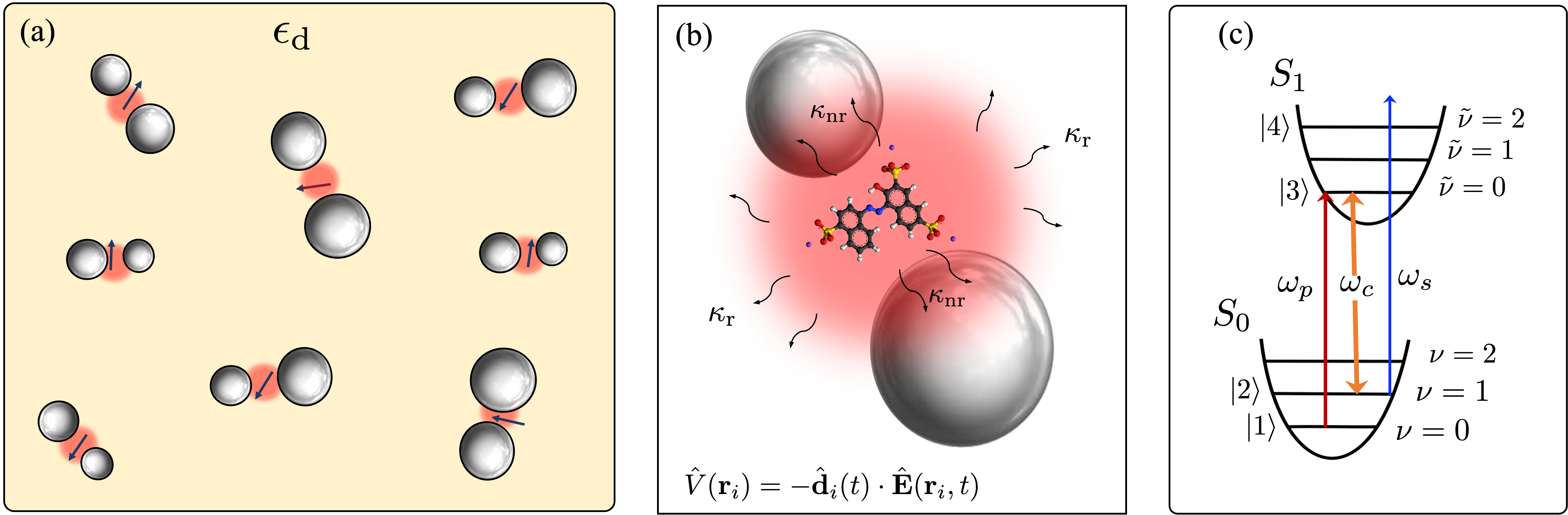}
\caption{{\bf Picocavity ensemble as an effective nonlinear medium}. (a) Ensemble of single-molecule picocavities as independent inclusions in a medium with dielectric constant $\epsilon_{\rm d}$. Each picocavity is represented by a local light-matter Hamiltonian $\hat H_i(t)$, arrows represent molecular dipoles. (b)  Dipolar coupling between a molecular dipole $\hat{\mathbf{d}}_i(t)$ and the local field $\hat{\mathbf{E}}(\r_i,t)$ of the $i$-th picocavity. The cavity field decays radiatively at rate $\kappa_{\rm r}$ and non-radiatively at rate $\kappa_{\rm nr}$. (c) Displaced oscillator model for the electronic ground ($S_0$) and first excited ($S_1$) electronic states molecular dipole. The local picocavity field drives the vibronic coherence between the $\nu=1$ and $\tilde \nu=0$ states at frequency $\omega_c$. The zero-phonon line is driven by a weak laser probe at frequency  $\omega_p$, and a signal laser drives a hot-band vibronic coherence between  $\nu=1$ and $\tilde \nu=2$, at frequency $\omega_s$.}
\label{fig:effective medium}
\end{figure*}

The phase-changing plasmonic metamaterial studied in this work is illustrated in Fig. \ref{fig:metasurface}. We consider an inhomogeneous layer composed of an isotropic ensemble of plasmonic nanoparticle inclusions dispersed in a background thin film with dielectric constant $\epsilon_{\rm d}$. Such substrates are routinely used in plasmon-enhanced molecular spectroscopy \cite{Lassiter2013,Esashika:19,Jais2011,Babonneau2018}. In our case, we assume that the nanoparticles self-assemble as dimer picocavities with a single organic molecule embedded in the gap region, where strong light-matter interaction occurs locally within the electric-dipole approximation (see Fig. \ref{fig:effective medium}b). Spherical nanoparticle dimers are common \cite{Esashika:19}, and single-molecule picocavities with a variety of geometries and material compositions can be produced \cite{Benz2016,Carnegie2018,Chikkaraddy:2016aa,Chikkaraddy2018}.

We estimate the impact of the ensemble of single-molecule picocavities on the effective refractive index of the metamaterial. Due to the absorptive and dispersive character of molecular picocavities, the metamaterial index depends on frequency. In particular, we focus on the phase response of the system at the frequency of a weak probe field $\omega_p$, which is resonant to the peak absorption frequency of the molecules, but is far-detuned from the mean gap resonance frequency of the picocavities $\omega_c$. We consider molecules with large vibronic coupling such that the excited state reorganization energy is comparable with the intramolecular vibration frequency ($\sim 0.2\, {\rm eV}$ \cite{Spano2010}). This ensures that molecules cannot strongly emit light at $\omega_p$, so the external probe laser is the relevant field source of the problem.

\subsection{Effective Index of the Metamaterial}
\label{sec:effective index}

\notablue{Effective medium theories are used for evaluating the 
average macroscopic electromagnetic response of a spatially inhomogeneous medium \cite{Sihvola-book}. For plasmonic nanoparticle ensembles, several techniques have been used for computing macroscopic dielectric response (effective index) \cite{ASPNES2011}, based on the well-known Maxwell Garnett theory \cite{Markel2016,Czajkowski2018}. We follow an alternative effective medium methodology developed in Refs. \cite{Lifshitz1,Lifshitz2}, which has been applied to describe  inhomogeneous metals \cite{Kaganova2003}, polycrystal dielectrics \cite{Kaganova1995} and  polaritons in organic microcavities \cite{LITINSKAIA2000}. The dynamical variables of the problem are written as the sum of a macroscopic average and a local fluctuation, for which coupled equations of motion can be derived and solved. }

We consider a dilute ensemble of $N$ {independent} single-molecule picocavities isotropically distributed over a background medium with dielectric constant $\epsilon_{\rm d}$, as illustrated in Fig. \ref{fig:effective medium}a. \notablue{The average picocavity size ($\sim 10$ nm) is much smaller than wavelength of the probe laser ($\sim 600-800$ nm), which is assumed to be the dominant electromagnetic source of the problem. In other words, inelastically scattered light from  plasmonic nanoparticles as well as fluorescence and Raman scattering from  organic molecules are negligibly weak at the probe frequency in comparison with the driving laser. }

We start with the wave equation 
\begin{equation}\label{eq:local wave eq1}
\nabla^2\E(\r,t)- \frac{1}{\epsilon_0c^2}\frac{\partial^2}{\partial t^2}\D_{\rm d}(\r,t) = \frac{1}{\epsilon_0c^2}\frac{\partial^2}{\partial t^2}\P(\r,t),
\end{equation}
where $\D_{\rm d}(\r,t)=\epsilon_0\epsilon_{\rm d}\E(\r,t)$ is the displacement field due to the background dielectric, and $\P(\r,t)$ is the polarization density due to picocavities with embedded dipoles, assumed to be large near the picocavities and vanishing in between. Computing this quantity from first principles is demanding, as it contains coupled charge density contributions from both the nanostructure and the embedded molecules \cite{Fojt2021}. \notablue{Field enhancement factors due to interfaces can also in principle be obtained by solving Eq.~(\ref{eq:local wave eq1}) numerically given a distribution of particle geometries and compositions. However, we focus our analysis on the generalities of the coupled light-matter response at a particular frequency (probe) and defer the exact evaluation of the spatially-dependent local fields for future work. }

We write Eq. (\ref{eq:local wave eq1}) in the Fourier domain by expanding the electric field and polarization into components at discrete frequencies $\omega_n$, to read
\begin{eqnarray}\label{eq:}
\nabla^2\E(\r,\omega_n) +\frac{\epsilon_{\rm d}\omega_n^2}{c^2}\,\mathbf{E}(\r,\omega_n) &=& -\frac{\omega_n^2}{\epsilon_0c^2}\mathbf{P}(\r,\omega_n).
\end{eqnarray}

For a weak probe field, the polarization at  $\omega_n\equiv\omega_p$ is expanded up to linear terms in the probe field, $E(\r)=\E_p(\r)$, as
\begin{equation}\label{eq:local P probe}
\P(\r,\omega_p) =\epsilon_0\chi(\omega_p,\omega_c, \omega_s, \E_c(\r),\E_s(\r))\cdot \E_p(\r,\omega_p),
\end{equation}
where the susceptibility $\chi\equiv\chi(\omega,\r)$ encodes the response at $\omega_p$ of the picocavities containing molecular dipoles. 

In general, the susceptibility in Eq. (\ref{eq:local P probe}) depends on the picocavity frequency $\omega_c$ and the local cavity fields strength, $\E_c(\r)$. We also anticipate a dependence of the effective susceptibility on the frequency $\omega_s$ and amplitude $\E_s(\r)$ of a signal laser field that is introduced in Sec. \ref{sec:level scheme} as an additional optical knob in the problem. In Sec. \ref{sec:homogeneous AT} we derive explicit expressions for $\chi$ using a quantum mechanical model for light-matter coupling in a picocavity.

Following Refs. \cite{Lifshitz1,Lifshitz2,Kaganova2003,Kaganova1995,LITINSKAIA2000}, we write the probe field and susceptibility as ($\omega \approx \omega_p$ is always assumed)
\begin{eqnarray}
\E_p(\r)&=&\langle \E_p(\r)\rangle + \delta\E_p(\r),\label{eq:local Ep1}\\
\chi(\r) & =& \langle \chi\rangle + \delta \chi(\r).\label{eq:local chi1}
\end{eqnarray}
where $\langle \E_p(\r)\rangle$ is the average probe field that propagates according to the effective index $n(\omega_p)$, and $\delta\E_p(\r)$ is a local field fluctuation, satisfying $\langle \delta \E_p(r)\rangle =0$. We expect that $\delta \E_p(\r)$ are non-zero only in the immediate vicinity of the picocavities. 

Similarly, the susceptibility is partitioned as a sum of a uniform average, $\langle \chi\rangle$, and a local fluctuation $ \delta \chi(\r)$, with $\langle \delta \chi(\r)\rangle=0$. Inserting Eqs. (\ref{eq:local P probe}), (\ref{eq:local Ep1}), and (\ref{eq:local chi1}) in Eq. (\ref{eq:local wave eq1}), we can derive coupled equations for the field and susceptibility averages and their fluctuations. We solve these equations in Appendix \ref{app:effective index} using perturbation theory, with the small parameter being the ratio of the typical inclusion scale to the wavelength of the electromagnetic wave in the medium. We find that the effective homogeneous index of the metamaterial at the probe frequency splits into three terms, 
\begin{equation}\label{eq:n full}
n^2(\omega_p)= \epsilon_{\rm d} + \langle \chi(\omega_p)\rangle + \Delta \epsilon_p,
\end{equation}
where $\langle \chi \rangle = (N/V)\chi_0$, with $\chi_0$ being the susceptibility of a single inclusion, is the average susceptibility, and
\begin{equation}\label{eq:delta epsilon}
\Delta \epsilon_p(\omega_p) = \frac{\omega_p^2}{c^2\epsilon_0^2} \int \frac{d{\bf q}\ K(|{\bf k}-{\bf q}|)}{\frac{\omega_p^2}{c^2}[\epsilon_\text{d} + \langle \chi \rangle]-q^2}
\end{equation}
is a correction to the homogeneous index, which takes into account spatial correlations between local fluctuations of the electric field and polarization, as captured by the Fourier transform of the spatial correlator $K(|\r_1-\r_2|) = \langle \delta\chi(\r_1) \delta\chi(\r_2)\rangle$.

Not going into detailed analysis of $\Delta\epsilon_p(\omega_p)$, which can be done explicitly for specific experimental configurations of metamaterial, in Appendix A we argue that the correlator $K(r)$ must vanish at the scales of the order of the inclusion size $r_0\sim$~10~nm; this simply reflects the assumption that the picocavities are independent. Assuming $K(r)\propto\exp[-r^2/(2r_0^2)]$ yields $K(q)\propto \exp[-q^2r_0^2/2]$, which is vanishingly small as long as $\lambda_p = 2\pi/q \gg r_0$. Having $\lambda_p\sim~700$~nm, below we neglect this correction, and set $n^2(\omega_p)= \epsilon_{\rm d} + \langle \chi(\omega_p)\rangle$.




\subsection{Global Phase Shift via Local Strong Coupling}
\label{sec:phase shifts}

In what follows we focus on the contribution of the ensemble of disordered single-molecule picocavities to the variation of the refractive index
$\Delta n(\omega_p)=\langle \chi(\omega_p)\rangle$, relative to a picocavity-free layer with dielectric constant $\epsilon_{\rm d}$. Detecting refractive index variations is  standard  in plasmonic sensing \cite{Xu2019-sensing}.

Due to the presence of picocavities, a probe wave that propagates in the metamaterial over a distance $L$ is phase shifted by
\begin{equation}\label{eq:relative phase}
 \Delta \Phi_L  \equiv {\rm Re}\{\Delta n(\omega_p)\}\omega_p L/c,
\end{equation}
relative to propagation in the pure background dielectric. Since ${\rm Im}\{\epsilon_{\rm d}\}\rightarrow 0$ and $|\langle \chi\rangle|\ll1$, we have that $\Delta n(\omega_p) \approx {\rm Re}\langle \chi(\omega_p)\rangle/2 $. The  susceptibility function scales as $\langle \chi\rangle \sim ({N}/{V}){|d_{eg}|^2}/{2\epsilon_0\hbar}$, with $N/V$ being the number density of single-molecule picocavities and $|d_{eg}|^2$ being the dipole moment for the transition induced at the probe frequency ($d_{eg}=3.8\, {\rm D}$ in Ref. \cite{Chikkaraddy2018}). Large phase variations $\Delta\Phi_L/{\Phi_L}$ of a few percent relative to the cavity-free background can be achieved with number densities $N/V\sim 10\;\mu{\rm m}^{-3}$, which is consistent with our dilute regime assumptions and experiments \cite{Papaioannou2016,Chikkaraddy2018}.

\section{Intracavity light-matter coupling scheme}
\label{sec:level scheme}

\begin{table*}[t]
\begin{tabular}{l | c | c}\hline\hline
{\rm Dissipative Process} &  {\rm Lindblad operator} ($\hat L_\alpha$) & {\rm Timescales}\\\hline
{\rm Cavity photon leakage} ($\kappa$) & $ \sqrt{\kappa}\hat a$ \,&  $\sim 10-10^2$ {\rm fs} \\
{\rm Intramolecular vibrational relaxation in} $S_0$ ($\gamma_{\rm v}$) &$\sqrt{\gamma_{\rm v}}\,\ket{1}\bra{2}$ & $\sim 1$ ps\\
{\rm Intramolecular vibrational relaxation in} $S_1$ ($\gamma'_{\rm v}$) &$\sqrt{\gamma'_{\rm v}}\,\ket{3}\bra{4}$ & $\sim 1$ ps\\
{\rm Dephasing of zero-phonon resonance} ($\gamma_{e}$)& $\sqrt{\gamma_{e}}\,\ket{1}\bra{3}$ & $\sim 1-10^3$ ps\\
\hline
\hline
\end{tabular}
\caption{Description and notation of the incoherent channels considered in this work (left), the associated Lindblad operators in the bare basis (center), and the corresponding decoherence timescales (right).}
\label{tab:lindblad}
\end{table*}

Optical phase variations induced by the intracavity molecules depend on the dynamics of the internal molecular coherences, which determine the frequency dependence of $\langle \chi(\omega_p)\rangle $. We study this dynamics for molecules within the lowest electronic potentials $S_0$ and $S_1$, as illustrated in Fig. \ref{fig:effective medium}c. The ground ($\nu=0$) and first excited ($\nu=1$) vibrational levels in $S_0$ are coupled to the lowest vibrational level ($\tilde \nu=0$) in $S_1$, by the probe and cavity fields, respectively. The Huang-Rhys factor \cite{Spano2011} in $S_1$ is large enough to give a sizable oscillator strength for the $\nu=0\leftrightarrow \tilde \nu=1$ vibronic sideband. The vibrational frequency $\omega_{\rm v}$ is assumed to exceed $k_{\rm b}T/\hbar$ at room temperature, as is typical with vinyl stretching modes ($\omega_{\rm v}\approx 0.18$ eV) \cite{Baranov2017}.
The picocavity frequency is set to $\omega_c=\omega_{0\tilde 0} -\omega_{\rm v}$, with $\omega_{0\tilde 0}$ being the $0\rightarrow \tilde0$ vibronic absorption frequency. The cavity detuning from the $\omega_{0\tilde 0}$ resonance ensures that the cavity field preferentially drives the $1-\tilde 0$ transition, and in particular prevents driving population out of the ground vibrational state ($\nu=0$) in the absence of the probe.

In addition to the probe field at frequency $\omega_p$ and the picocavity field at $\omega_c$, we introduce in Fig. \ref{fig:effective medium}c an additional classical {\it signal field} at frequency $\omega_s$. The signal field drives the hot vibronic absorption band $\nu=1\rightarrow \tilde \nu=2$ off-resonantly. We show later that this signal field can be used as an optical switch, for controlling the molecular susceptibility at the probe frequency. In summary, we have the frequency hierarchy $\omega_c<\omega_p<\omega_s$.

In order to compute $\langle \chi(\omega_p) \rangle$, we label the relevant molecular transitions as $\ket{1}\equiv \ket{\nu=0}$, $\ket{2}\equiv \ket{\nu=1}$, $\ket{3}\equiv \ket{\tilde \nu=0}$, and $\ket{4}\equiv \ket{\tilde \nu=2}$), according to the scheme in Fig. \ref{fig:effective medium}c, to write a picocavity Hamiltonian of the form (we use $\hbar \equiv 1$ throughout)
\begin{eqnarray}\label{eq:Hamiltonian}
\hat{\mathcal{H}}&=&\omega_c\,\hat a^\dagger \hat a+ \omega_{21}\ket{2}\bra{2}+\omega_{31}\ket{3}\bra{3}+\omega_{41}\ket{4}\bra{4}\nonumber\\
&&+g_c \ket{3}\bra{2}\hat a+\Omega_p \ket{3}\bra{1}{\rm e}^{-i\omega_p t} +  \Omega_s \ket{4}\bra{2}{\rm e}^{-i\omega_s t}\nonumber\\
&&+ {\rm H.c.}
\end{eqnarray}
where $\omega_{ij}=(E_i-E_j)/\hbar$ denote the molecular transition frequencies, $g_c$ is the picocavity vacuum Rabi frequency, $\Omega_p$ is the classical Rabi frequency of the probe field,  and $\Omega_s$ is the classical signal Rabi frequency. The bosonic cavity field operator is $\hat a$, and H.c. stays for Hermitian conjugation. Energy is given relative to the ground vibrational level (i.e., $E_1=0$). \notaviolet{Although the analysis reduces to an effective four-level system, the displaced oscillator picture for the $S_0$ and $S_1$ manifolds is helpful for anticipating potential issues in an experimental implementation of the scheme. It also serves as the basis of further studies that take energy transport dynamics into account \cite{Kessing2021}.}

For a quantized picocavity in the low-excitation manifold, the molecular basis should be supplemented with  Fock states $|n_c\rangle$, to give the dressed  basis: $\ket{\tilde 1}\equiv \ket{1;0_c}$, $\ket{\tilde 2}\equiv \ket{2;1_c}$,  $\ket{\tilde 3}\equiv \ket{3;0_c}$, and $\ket{\tilde 4}\equiv \ket{4;1_c}$. In the dressed-state picture, the probe field $\Omega_p$  drives the transition $\ket{\tilde 1}\leftrightarrow\ket{\tilde 3}$ within the vacuum manifold, and the signal field $\Omega_s$ drives the transition $\ket{\tilde 2}\leftrightarrow\ket{\tilde 4}$ within the one-photon manifold. The picocavity field admixes the vacuum and one-photon states $\ket{\tilde 2}$ and $\ket{\tilde 3}$. The coupling of the quantized picocavity field with other molecular transitions is neglected. 

We model the evolution of the reduced density matrix $\hat \rho(t)$ for an individual picocavity with a Lindblad quantum master equation of the form \cite{Litinskaya2019}
\begin{equation}\label{eq:Lindblad}
\frac{d}{dt}\hat\rho = -i[\hat{\mathcal{H}}, \hat \rho]+\sum_\alpha \hat L_\alpha \hat \rho\hat L_\alpha^\dagger - \frac{1}{2}(\hat L_\alpha^\dagger\hat L_\alpha \hat \rho +\hat \rho  \hat L_\alpha^\dagger\hat L_\alpha),
\end{equation}
where $\hat{\mathcal{H}}$ is given in Eq. (\ref{eq:Hamiltonian}), and $\hat L_\alpha$ is the Lindblad operator associated with the $\alpha$-th dissipative channel. In Table \ref{tab:lindblad}, we list the Lindblad operators used in this work, and the associated dephasing times in the bare basis. We use a notation in which the rate $\gamma_{ij}$ describes the decay of the off-diagonal element $\rho_{ij}\equiv \langle i|\hat \rho|\ j \rangle$, in the dressed basis. For example, vibrational relaxation from $\nu=1$ (state $\ket{2}$) to $\nu=0$ (state $\ket{1}$) in $S_0$ occurs at the rate $\gamma_{\rm v}\sim 1\, {\rm ps}^{-1}$. On the other hand,  the bare picocavity photon lifetime is $\kappa^{-1}\sim 10-100 \,{\rm fs}$ \cite{Baranov2017}. Therefore, the combined decay rate of the dressed Raman coherence $\langle \tilde 1|\hat \rho|\tilde 2\rangle$ is $\gamma_{21}=\gamma_{\rm v}/2+\kappa/2$, as $\ket{\tilde 2}$ has a single-photon character. The lifetime of the excited electronic state $S_1$ is $\gamma_{31}\equiv \gamma_e$, given by either fluorescence or internal conversion. We set  $\gamma_{41}\equiv \gamma_{31}$ throughout.
\\
\\
\\

\section{Homogeneous Autler-Townes response}
\label{sec:homogeneous AT}

Starting from the master equation in Eq. (\ref{eq:Lindblad}), we follow Ref. \cite{Litinskaya2019} and derive a general expression for susceptibility for a homogeneous ensemble of $N$ independent molecules in identical single-molecule picocavities, subject to classical driving by the probe and signal fields. \notablue{We reproduce here the final expression, given by}
\begin{widetext}
\begin{equation}\label{eq:chi generic}
\chi(\omega_p) =\left(\frac{N}{V}\right)\frac{|d_{13}|^2}{\epsilon_0\hbar}\times \frac{ [\Delta_{21} + i \gamma_{21}] [\Delta_{41} + i \gamma_{41}] - \Omega_s^2}{
[\Delta_{31} + i \gamma_{31}] \left([\Delta_{21} + i \gamma_{21}][\Delta_{41} + i \gamma_{41}] - \Omega_s^2\right) - g_c^2 [\Delta_{41} + i \gamma_{41}]},
\end{equation}
\end{widetext}
\notablue{and refer readers to  Appendix \ref{app:chi} for the technical details of the derivation. } In Eq. (\ref{eq:chi generic}) we denote $|d_{31}|^2$ as the 0-$\tilde 0$ oscillator strength, $\Delta_{31} \equiv  \omega_p - \omega_{31}$ is the probe detuning, $\Delta_{21} \equiv \omega_p-\omega_c-\omega_{21}=\Delta_{31} - \Delta_c$ is the two-photon Raman detuning. Introducing the cavity detuning $\Delta_{c}\equiv \omega_{c}-\omega_{32}$ and the signal detuning $\Delta_s\equiv \omega_s-\omega_{42}$, we can write $\Delta_{41} \equiv \Delta_{31} - \Delta_c + \Delta_s$. The ensemble susceptibility in Eq.~(\ref{eq:chi generic}) involves only the single-molecule coupling strength $g_c$, highlighting the local character of the cavity-induced nonlinearity. 

\notaviolet{Although we focus on the polarization component at the probe frequency to derive Eq. (\ref{eq:chi generic}),  steady-state solutions for other polarization components at the cavity and signal field frequencies can also be derived from the Lindblad quantum master equation, resulting in a more general coupled mode theory that would describe the coherent interaction between different field components mediated by the material degrees of freedom \cite{Phillips2011,Engelhardt2021prl}. Solving this more general problem is beyond the current scope of this work. }

\begin{figure}[bt]
\includegraphics[width=0.40\textwidth]{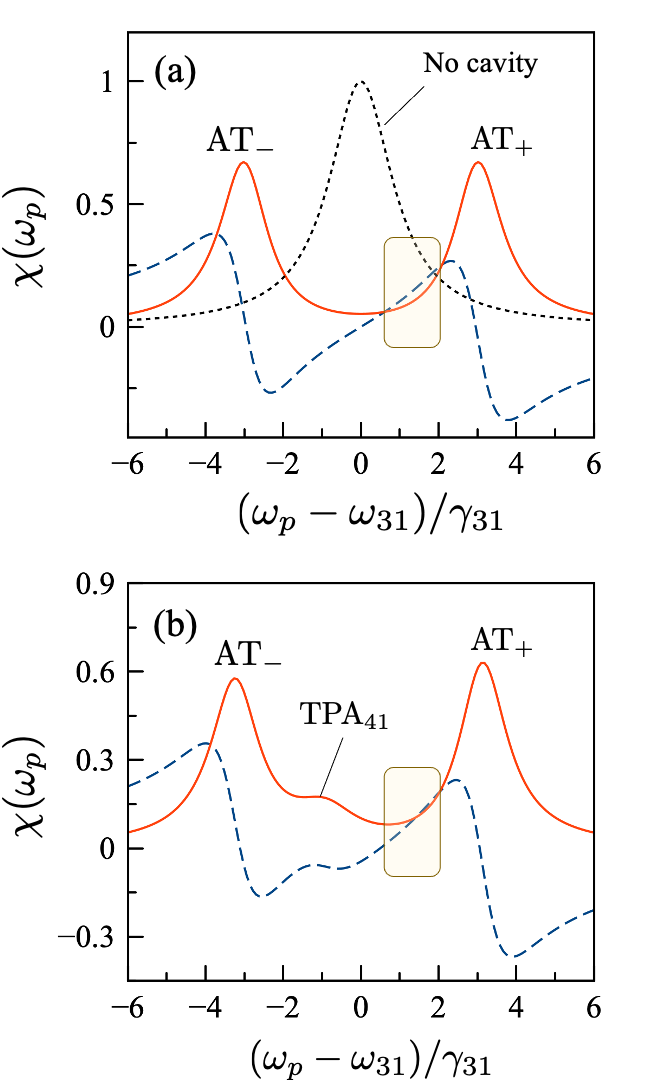}
\caption{{\bf Disorder-free molecular susceptibility}. (a) Absorptive (solid line) and dispersive (dashed line) components of the homogeneous susceptibility $\chi(\omega_p)$ near the bare probe resonance (dotted line), in the absence of a signal field. The Autler-Townes (AT) doublet is shown. System parameters are $(g_c,\gamma_{21},\gamma_{42})=(3.0,0.5,1.0)$, in units of the  homogeneous linewidth $\gamma_{31}$. (b) Absorptive and dispersive AT response  with the same parameters as in panel (a), but in the presence of a signal field with Rabi frequency $\Omega_s=\gamma_{31}$, blue detuned by $\Delta_s=\gamma_{31}$ from the $2\rightarrow 4$ transition. The AT doublet and a two-photon absorption (TPA$_{41}$) peak are shown. The shaded area shows a frequency region where, in the absence of the signal field, dispersion overcomes absorption. For smaller signal field amplitudes, the transparency window at this region remains open.}
\label{fig:AT response}
\end{figure}

In Fig. \ref{fig:AT response}, we plot the absorptive and dispersive parts of the disorder-free susceptibility $\chi(\omega_p)$ around the bare 0-$\tilde 0$ absorption resonance, under conditions of strong intracavity coupling. In Fig. \ref{fig:AT response}a, we show the system response without the signal field ($\Omega_s=0$). The absorptive response shows two Autler-Townes (AT) peaks at $\omega_p\approx \omega_{31}\pm g_c$. The doublet  opens a broad semi-transparent window (solid line), due to cavity-induced AT splitting of the dressed states $\ket{\tilde 3}$ and $\ket{\tilde 2}$.
The width of the Autler-Townes transparency window $\Gamma_{\rm AT}$ can be defined by $A(\Gamma_{\rm AT}/2)=A(g_c)/2$, with $A(\omega)\equiv {\rm Im}{\langle \chi(\omega)\rangle}$. For a homogeneous system, $\Gamma_{\rm AT}$ scales linearly with the cavity coupling $g_c$ \cite{Litinskaya2019}. On the other hand, the amount of residual absorption at the bare probe resonance ($\Delta_p=0$) can  be shown to scale with the ratio $\gamma_{21}/\gamma_{31}$ \cite{Fleischhauer:2005}.

Figure \ref{fig:AT response}a shows that at the center of the AT doublet, the probe field experiences normal dispersion (dashed line), in contrast to the 
``slow-light" dispersion expected for an interference-based transparency window \cite{Fleischhauer:2005}. The plot also shows  a relatively broad region within the AT transparency window (order $\gamma_{31}$ in frequency) in which probe dispersion overcomes absorption, i.e., ${\rm Re}\chi(\omega_p)>{\rm Im}\chi(\omega_p)$. Clearly this condition always holds away from absorptive resonances (e.g., $|\Delta_p|>6\gamma_{31}$ in Fig. \ref{fig:AT response}a). However, the ability of introducing a dispersive response in a frequency range that is otherwise opaque is a key resource for optical switching (see below).

In Fig. \ref{fig:AT response}b, we show the susceptibility for the same conditions as in Fig. \ref{fig:AT response}a, but now in the presence of a strong signal field ($\Omega_s=\gamma_{31}$), detuned from  the $2\rightarrow4$ transition frequency ($\Delta_s=\gamma_{42}$). The positions of the peaks ${\rm AT}_{\pm}$ remain practically unchanged, and the probe response continues to exhibit the same AT doublet. However, now a two-photon absorption resonance (TPA$_{41}$) destroys the transparency window: we see a broad background with a peak at the TPA$_{41}$ resonance condition, $\Delta_p-\Delta_c+\Delta_s=0$.
For $\Delta_c=0$ and $\Delta_s=\gamma_{31}$, the resonance  occurs at $\Delta_p= -\gamma_{31}$. This TPA channel can be understood as a result of the mixing between the dressed states $\ket{\tilde 2}$ and $\ket{\tilde 3}$, mediated by an intracavity vacuum that acts as an effective doorway mechanism for the 
$(\omega_p,\omega_s)$ process:
$$\ket{1}\ket{0_c}\xrightarrow{\omega_p}(\ket{3}\ket{0_c}\leftrightarrow \ket{2}\ket{1_c})\xrightarrow{\omega_s} \ket{4}\ket{1_c}.$$  Note that the classical probe and signal fields do not change the intracavity photon number. This emerging TPA channel effectively modulates the dispersive response of the probe field within the AT window, by making absorption comparable to the dispersion (Fig. \ref{fig:AT response}b, shaded area). As we discuss below, at lower signal beam intensities the detrimental effect of the signal field can be small enough to preserve its dispersive properties in spite of increased absorption.  \notaviolet{Additional nonlinear absorption channels such as hot absorption $1-\tilde 1$ can be suppressed, either by controlling the probe field intensity or selecting molecular vibrations with different fundamental frequencies in $S_0$ and $S_1$ (e.g., azobenzene \cite{Casellas2016}). }

\section{Disorder-averaged Autler-Townes response}
\label{sec:inhomogeneous AT}

\begin{figure}[t]
\includegraphics[width=0.30\textwidth]{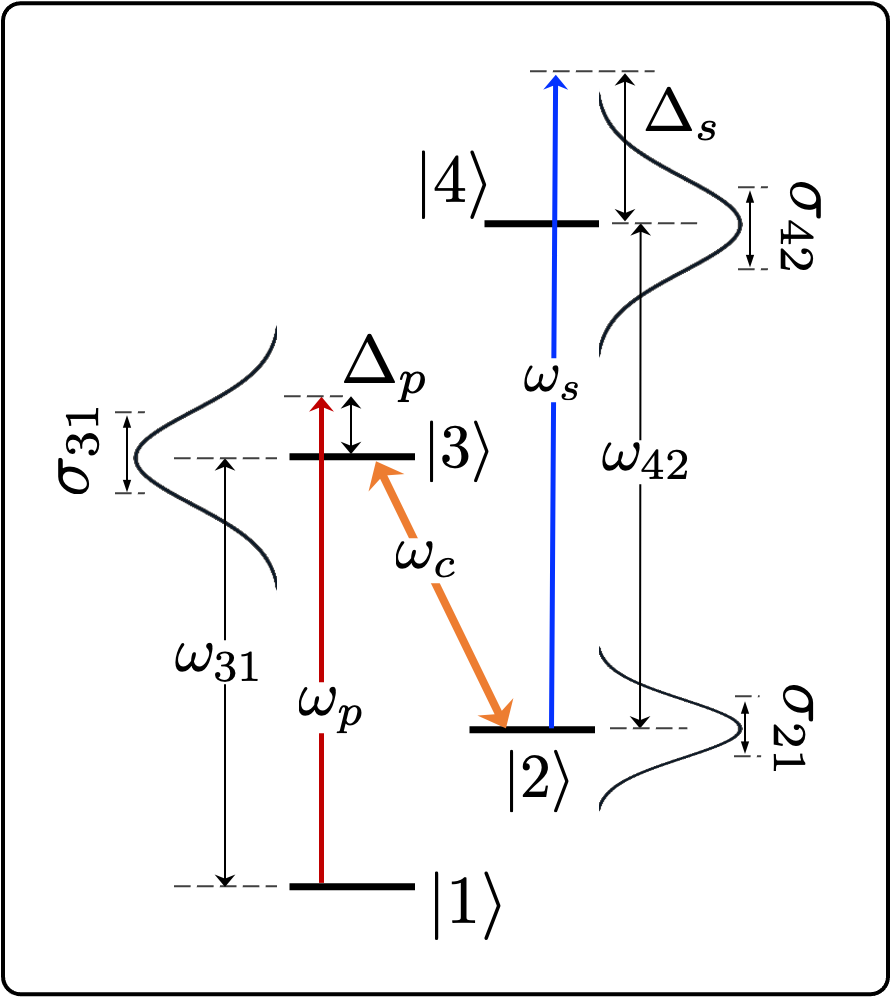}
\caption{{\bf Inhomogeneous broadening scheme}. The probe absorption band at $\omega_{31}$ is inhomogeneously broadened due to static energy disorder by $\sigma_{31}$ (FWHM$\approx 2.355 \sigma_{31}$). The ground state vibrational band at $\omega_{21}$ is broadened by $\sigma_{21}$, and the inhomogeneous width of the hot-band absorption band is $\sigma_{42}$. The probe and signal fields are detuned from the peak frequencies by $\Delta_p$ and $\Delta_s$, respectively. The $\rho_{32}$ vibronic coherence is driven by an ensemble of cavity fields at $\omega_c$, leading to a Gaussian distribution of Rabi frequencies with standard deviation $\sigma_{g_c}$.}
\label{fig:disorder scheme}
\end{figure}

\subsection{Rabi and energy disorder}

In this section, we study the lineshape of the AT transparency window under two types of structural disorder: random intracavity volumes and random molecular transition frequencies. The first arises from the distribution of gap volumes $V_{\rm g}$ in the picocavity ensemble. The volume distribution is assumed to be Gaussian, with mean value $\langle V_g\rangle$ and standard deviation $\sigma_{V_{g}}$. Given that $g_c= f/\sqrt{V_g}$ \cite{Maier2006}, with $f$ a constant, the distribution of Rabi frequencies $g_c$ is also a Gaussian with mean value $\langle g_c\rangle=f/\sqrt{\langle V_g\rangle }$ and variance
%
$\sigma_{g_c}^2=f^2\sigma_{V_g}/2\langle V_g\rangle^{3/2}$, where only leading terms in the small parameter $\sigma_{V_g}/\langle V_g \rangle$ are kept.

In Fig. \ref{fig:disorder scheme}, we illustrate the impact of Gaussian energy disorder of the molecular transition frequencies. The inhomogeneous width of the cavity-free probe absorption band is  $\sigma_{31}$ (${\rm FWHM}\approx 2.4 \,\sigma_{31}$). The ground state vibrational band $\nu=0\rightarrow \nu=1$  has inhomogeneous width $ \sigma_{21}$, and width of the hot-band absorption   $\nu=1\rightarrow \tilde \nu=2$ is $\sigma_{42}$. Since pure vibrational linewidths ($\sigma\sim 4$ meV \cite{Pollard2014}) are much smaller than typical vibronic linewidths ($\sigma\sim 100$ meV \cite{Spano2011}), we have  $\sigma_{21}/\sigma_{31}\ll 1$ and $\sigma_{42}\sim \sigma_{31}$.

\begin{figure}[t]
\includegraphics[width=0.4\textwidth]{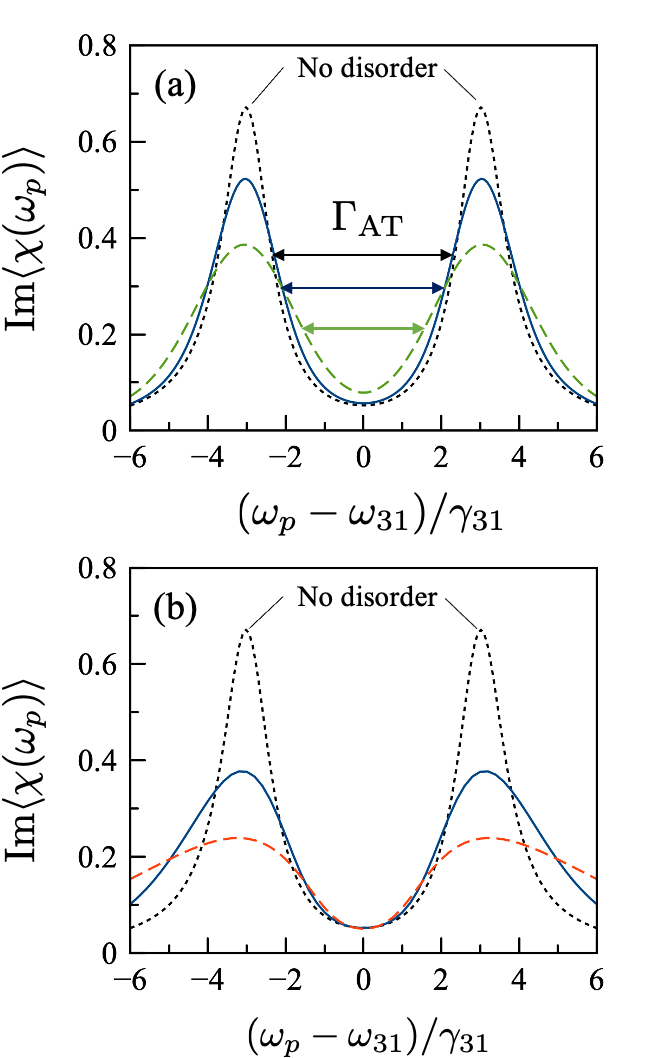}
\caption{{\bf Autler-Townes doublet with inhomogeneous broadening}. (a) Probe absorption lineshape for a Gaussian distribution of Rabi couplings with mean value $\langle g_c\rangle =3\gamma_{31}$, and variable standard deviation $\sigma_{g_c}/\gamma_{31}=0$ (dotted line), 0.5 (solid line), and 1.0 (dashed line). No signal field is present and molecular transitions are homogeneously broadened, with $(\gamma_{21},\gamma_{42})=(0.5,1.0)$, in units of $\gamma_{31}$. The Autler-Townes width $\Gamma_{\rm AT}$ is highlighted. (b) Absorption lineshape for a narrow Rabi frequency distribution with $(\langle g_c\rangle,\sigma_{g_c})=(3.0,0.01)$, in units of $\gamma_{31}$, and inhomogeneously broadened molecular levels with $\sigma_{31}/\gamma_{31}=0$ (dotted line), 2.0 (solid line), and 4.0 (dashed line).  In both panels, we set $\Delta_c=0$.}
\label{fig:disorder doublet}
\end{figure}

In Fig. \ref{fig:disorder doublet}a, we plot the AT absorption doublet for an ensemble with a Gaussian distribution of Rabi frequencies, but  otherwise homogeneous, in the absence of a signal field ($\Omega_s=0$). In comparison with the fully homogeneous response from the previous section, the doublet lineshape remains largely unaltered even for broad distributions with $\sigma_{g_c}\approx \langle g_c\rangle$. This is reminiscent of the weak dependence of the cavity response on the distribution of dipole moment orientations found in Ref. \cite{Litinskaya2019}. Therefore we neglect both orientational and mode volume disorder in what follows.

In Fig. \ref{fig:disorder doublet}b, the AT lineshape is shown for a fixed Rabi frequency ($\sigma_{g_c}\rightarrow 0$), but the molecular levels are inhomogeneously broadened, and the response is numerically averaged over a Gaussian distribution of energy levels. No signal field is applied. We see that the overall doublet shape of the AT transparency window is insensitive to the increase of the vibronic linewidth $\sigma_{31}$. This is expected, as the AT transparency window is sustained by the cavity-induced Raman coherence $\rho_{21}=\langle 1,0_c|\hat \rho|2,1_c\rangle$, which is limited by the photon decay rate $\kappa$.

\begin{figure}[t]
\includegraphics[width=0.40\textwidth]{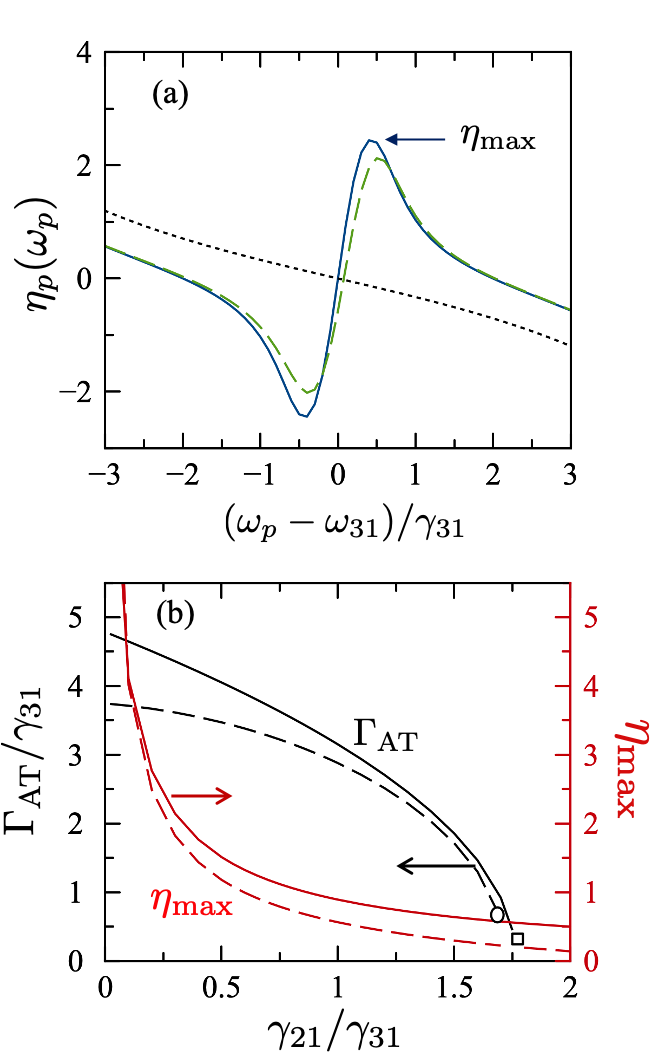}
\caption{{\bf Autler-Townes transparency in a disordered ensemble}. (a) Figure-of-merit $\eta_p$ for the refractive index variation within an Autler-Townes window with system parameters $(g_c, \sigma_{31},\sigma_{21})=(2.0,2.0,0.01)$, in units of $\gamma_{31}$ (numerical averaging). Curves are shown in the absence of the signal field (solid line), and in the presence of a signal field with $\Omega_s=0.6\gamma_{31}$ and $\Delta_s=1\,\gamma_{31}$ (dashed line). The maximum figure-of-merit $\eta_{\rm max}$ is highlighted. Cavity-free results for the same broadening parameters are also shown (dotted line). (b) Left axis shows the Autler-Townes transparency width $\Gamma_{\rm AT}$ as a function of  $\gamma_{21}$, for Lorentzian disorder (solid line) and Gaussian disorder (dashed line), with system parameters $(g_c, \sigma_{31},\sigma_{21},\Omega_s)=(3.0,2.0,0.01,0)$, in units of $\gamma_{31}$. The right axis shows the corresponding optimal figure-of-merit ratio $\eta_{\rm max}$ for Lorentzian disorder (solid line) and Gaussian disorder (dashed line). }
\label{fig:disorder scaling}
\end{figure}

\subsection{Theoretical limits for phase modulation}

We now study the feasibility of the proposed optical switch in disordered picocavity ensembles. In order for the relative phase shift $\Delta\Phi_L/\Phi_L$ in Eq. (\ref{eq:relative phase}) to be detectable, dispersion of the probe field should overcome absorptive losses. Therefore, we use the figure-of-merit
\begin{equation}\label{eq:eta definition}
 \eta_p (\omega_p)\equiv \frac{{\rm Re} \chi(\omega_p)}{{\rm Im} \chi(\omega_p)},
 \end{equation}
to quantify the theoretical performance of the metamaterial for phase modulation at the probe frequency. The global dispersive behavior of the medium (phase shifting) prevails over inherent molecular absorption losses when $\eta_p$ exceeds unity over a bandwidth $\gamma_{31}$. In a heterodyne setting that detects the interference of a transmitted probe field ${E}_p(\omega_p)$ with a reference beam, the ratio $\eta_p$ correlates with the fringe visibility. If the transmitted probe field is largely attenuated ($\eta_p\ll1$), interference with a reference field cannot be resolved.

To account for energy disorder, we average Eq. (\ref{eq:eta definition}) numerically using independent Gaussian distributions for each molecular transition frequency. We also estimate this average analytically using independent Lorentzian distributions for the molecular transition frequencies. As we prove in Appendix \ref{app:Lorentzian}, the Lorentzian averaging reduces to replacing $\gamma_{ij}$ in Eq. (\ref{eq:chi generic}) with $\Sigma_{ij} = \gamma_{ij}+\sigma_{ij}$ everywhere. These analytical results allow us to gain insight into the multiple parameters that determine the effective nonlinear probe response, and we confirm numerically that a more realistic Gaussian disorder gives the same trends for key observables as the Lorentzian disorder model. \notaviolet{The Lorenztian disorder technique has also been used in Refs. \cite{Vlaming2009,Litinskaya2019,Engelhardt2021} to simplify the average of system observables over a random distribution of Hamiltonian parameters. }

We can rewrite Eq.(\ref{eq:eta definition}) for the figure-of-merit by separating the real and imaginary parts of the susceptibility (\ref{eq:chi generic}). As a next step, we apply our technique of averaging over Lorentzian disorder, developed in Ref. \cite{Litinskaya2019} and described in detail in Appendix \ref{app:Lorentzian}. In the absence of a signal field ($\Omega_s=0$), the ratio $\eta_p$ as a function of probe detuning $\Delta_p$ for a resonant cavity ($\Delta_c=0$) becomes:
\begin{equation}\label{eq:eta analytic}
\eta_p(\Delta_p) =  \frac{g_c^2 \Delta_p-\Delta_p(\Delta_p^2 + \Sigma_{21}^2) }{\Sigma_{31}(\Delta_p^2 + \Sigma_{21}^2) + \Sigma_{21} g_c^2},
\end{equation}
where the parameters $\Sigma_{ij}\equiv \gamma_{ij}+\sigma_{ij}$ represent total decoherence rates, including  homogeneous ($\gamma_{ij}$) and inhomogeneous ($\sigma_{ij}$) contributions. For the numerical averaging over Gaussian disorder, we used this expression with $\Sigma_{ik}\to\gamma_{ik}$.

In a cavity-free scenario, Eq. (\ref{eq:eta analytic}) reduces to the linear scaling $\eta_p=-\Delta_p/\Sigma_{31}$. This linear dependence is reproduced also if the averaging over inhomogeneous broadening is carried out numerically using Gaussian disorder (shown as a dotted line in Fig. \ref{fig:disorder scaling}a). The cavity vacuum induces a deviation from this linear scaling. As numerical averaging shows (solid and dashed lines in Fig. \ref{fig:disorder scaling}a), the figure-of-merit increases and exhibits a maximum when the probe is slightly detuned from the center of the AT window, with the maximum value $\eta_\text{max}$ considerably exceeding 1. We can estimate the optimal detuning at which $\eta_\text{max}$ is reached using our analytical Lorentz-averaged model (\ref{eq:eta analytic}). We find:   
\begin{equation}
\Delta_{p, \text{optimal}} \approx g_c\sqrt{\frac{\Sigma_{21}}{\Sigma_{31}+3\Sigma_{21}}},
\end{equation}
where we dropped quadratic terms in $\Sigma_{21}/g_c$. For low-quality picocavities in strong coupling regime, we have $\gamma_{\rm v}\ll \kappa\lesssim \gamma_{31}<g_c$.

We now can estimate $\eta_\text{max} = \eta_p(\Delta_{p, \text{optimal}})$. In the framework of the Lorentzian disorder model, the maximum figure-of-merit $\eta_{\rm max}$ is:
\begin{equation}\label{eq:eta max}
\eta_{\rm max}= \frac{g_c}{\sqrt{\Sigma_{21}(\Sigma_{31}+3\Sigma_{21})}}\left(\frac{\Sigma_{31}+2\Sigma_{21}}{2\Sigma_{31}+3\Sigma_{21}}\right),
\end{equation}
where we have ignored terms that are second order in $\Sigma_{21}/g_c$. If $\Sigma_{21}$ is small, then $\eta_{\rm max}$ is large, which is the case we studied in Ref. \cite{Litinskaya2019}. 
Consider now an ensemble of lossy picocavities with a large Raman decoherence rate $\Sigma_{21}\sim \Sigma_{31}$. Equation (\ref{eq:eta max}) then predicts that cavity-mediated optical phase modulation within the AT transmission window can still be feasible, provided that the single-molecule Rabi coupling is strong enough. For example, we can achieve $\eta_{\rm max}\geq 1$ even for $\Sigma_{21}/\Sigma_{31}=1$, with Rabi couplings $g_c/\Sigma_{31} \geq 3.33$. For a representative zero-phonon linewidth $\Sigma_{31}\approx 50$ meV \cite{Schmid2013,Chikkaraddy2018}, this corresponds to $g_c\approx 167$ meV. Improving the quality factor of the picocavities ($\Sigma_{21}\sim \kappa$), such that the ratio $\Sigma_{21}/\Sigma_{31}$ decreases by a factor of two,  reduces the constraint on single-molecule coupling to $g_c\geq 98$ meV, for  phase modulation to be detectable. Single-molecule couplings of these magnitudes are within experimental reach \cite{Benz2016,Chikkaraddy:2016aa,Chikkaraddy2018,Wang2017}.

Let us now check if the Lorentz approximation is consistent with numerical Gaussian-based approach. For the system parameters used in Fig. \ref{fig:disorder scaling}a, the Lorentz disorder model predicts the maximum figure-of-merit $\eta_{\rm max}\approx 2.0 $ at $\Delta_p=0.55\,\gamma_{31}$. This detuning is only slightly higher than the value $0.41\,\gamma_{31}$, predicted by numerically averaging Eq. (\ref{eq:eta analytic}), with $\Sigma_{ik}\to\gamma_{ik}$, over independent Gaussian frequency distributions (Fig. \ref{fig:disorder scaling}a, solid line). In the presence of a signal field ($\Omega_s>0$), blue-detuned from the 4-2 resonance by $\Delta_s=\gamma_{13}$, the AT window lineshape becomes distorted (see Fig. \ref{fig:AT response}), and $\eta_{\rm max}$ decreases monotonically with increasing $\Omega_s$ (Fig. \ref{fig:disorder scaling}a, dashed line), as discussed in more detail in Section \ref{sec:signal}.

Complementary to the discussion of the frequency-dependent phase performance parameter $\eta_p(\omega_p)$, we can use the Lorentz disorder model to analyze the lineshape of the Autler-Townes window. As already mentioned, the position of the AT$_\pm$ doublet peaks is largely insensitive to disorder. However, transparency within the AT window is reduced as the quality of the vibrational Raman coherence $\rho_{21}$ degrades with increasing $\gamma_{21}$ and $\sigma_{21}$, which decreases the width $\Gamma_{\rm AT}$.

In the Lorentz disorder model, the AT width can be written as
\begin{equation}\label{eq:Gamma AT approx}
\Gamma_{\rm AT}\approx 2\sqrt{g_c^2-\sqrt{2}g_c(\Sigma_{31}+\Sigma_{21})},
\end{equation}
which shows that the AT window is formally closed when
$\,{g_c} \leq \sqrt{2}(\Sigma_{21}+\Sigma_{31})$.

Figure \ref{fig:disorder scaling}b shows that the Lorentz disorder model in general overestimates $\Gamma_{\rm AT}$ relative to Gaussian averaging. For the parameters in Fig. \ref{fig:disorder scaling}b, the AT window is predicted by Eq. (\ref{eq:Gamma AT approx})  to close for  $\gamma_{21}=1.78\,\gamma_{31}$, which agrees well with the value ($1.71\gamma_{31}$) obtained for Gaussian disorder. Figure \ref{fig:disorder scaling}b also shows that Eq. (\ref{eq:eta max}) correctly captures the scaling of $\eta_{\rm max}$ with the Raman decay rate $\gamma_{21}$, for narrow vibrational coherences with $\sigma_{21}\ll \sigma_{31}$.

We conclude that $\eta_p > 1$, which correspond to observable phase shifts, can be achieved in our metamaterial in spite of strong energy disorder and ultra-fast cavity photon decoherence, as long as the coupling constant $g_c$ is large enough.


\subsection{Controlling the probe phase shift with a \\ weak signal laser}\label{sec:signal}

\begin{figure}[t]
\includegraphics[width=0.45\textwidth]{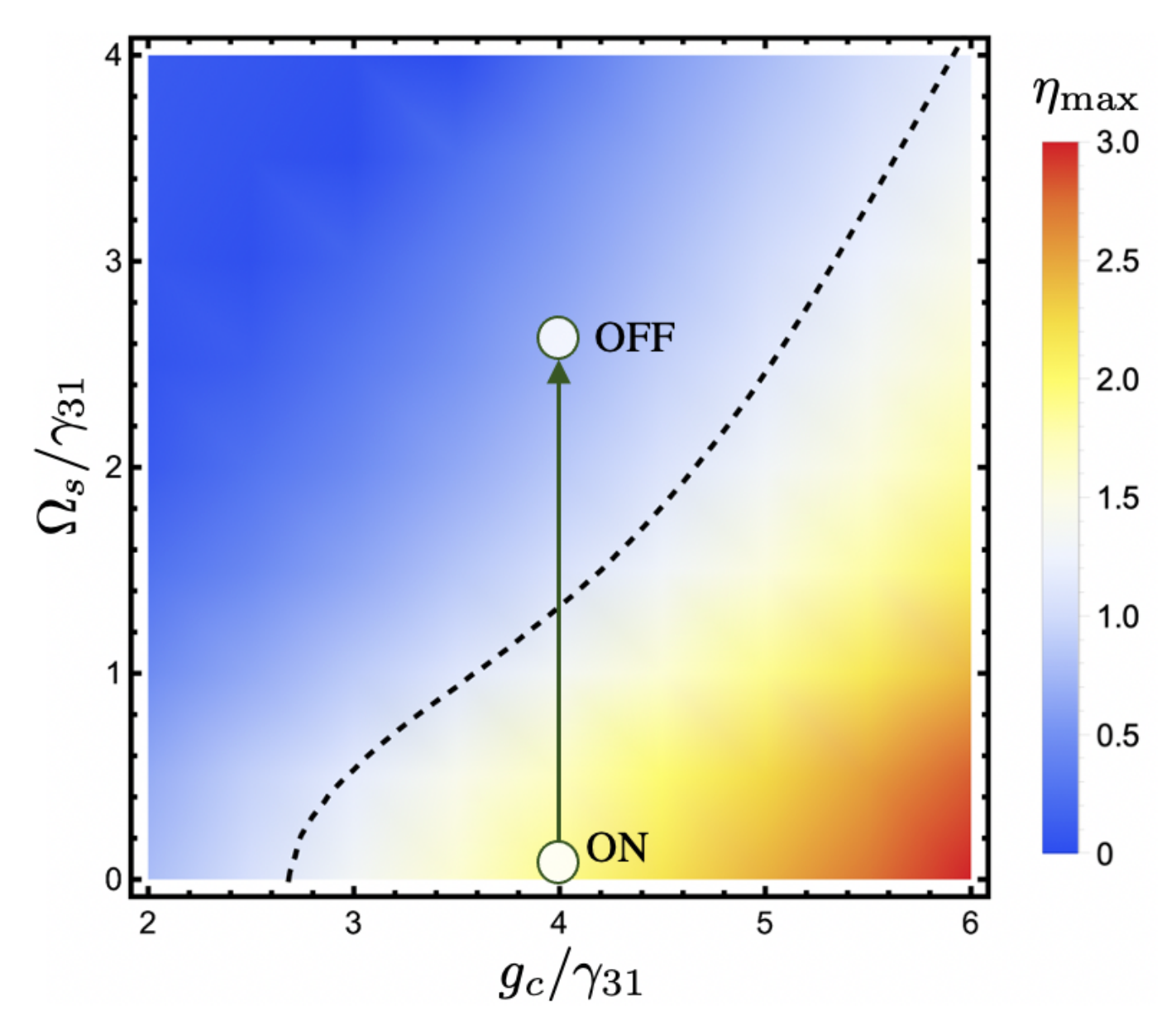}
\caption{{\bf Performance parameter for optical phase switching}. Optimal performance parameter $\eta_{\rm max}$ as a function of the cavity coupling strength $g_c$ and signal strength $\Omega_s$, in an Autler-Townes window corresponding to  $(\sigma_{31},\gamma_{21},\sigma_{21},\Delta_s)=(2.0,0.5, 0.01, 1.0)$, in units of $\gamma_{31}$. The dashed contour marks the detection limit $\eta_{\rm max}=1$, and two possible ON-OFF phase switch configurations are highlighted.}
\label{fig:switch diagram}
\end{figure}

Let us finally discuss the nonlinear interaction between the probe field at $\omega_p$ and an additional signal laser at $\omega_s$, mediated by the ensemble of single-molecule picocavities. In the presence of the signal field, we derive from Eq. (\ref{eq:chi generic}) a more general expression for $\eta_p$ than Eq. (\ref{eq:eta analytic}), which captures the dependence of $\chi_p$ on the signal frequency $\omega_s$ and its Rabi frequency $\Omega_s\propto |\E_s|$. We then average $\eta_p$ over independent fluctuations of the molecular transition frequencies $\omega_{31}, \omega_{32}$ and $ \omega_{42}$. Assuming the Lorentzian disorder model and applying the Lorentz averaging technique outlined in the Appendix \ref{app:Lorentzian}, we obtain a disorder-averaged expression for $\eta_p$ that reads~\cite{Litinskaya2019}

\begin{equation}\label{eq:eta_full}
\begin{array}{l}
\eta_p(\Delta_p) = \\
\\
\displaystyle \frac{g_c^2 (\Delta_p - \lambda_s \Delta_{41}) - \Delta_p [(\Delta_p - \lambda_s \Delta_{41})^2+(\Sigma_{21}+\lambda_s \Sigma_{41})^2]}{\Sigma_{31}[(\Delta_p - \lambda_s \Delta_{41})^2+(\Sigma_{21}+\lambda_s \Sigma_{41})^2]+ g_c^2(\Sigma_{21}+\lambda_s \Sigma_{41})},\\
\end{array}
\end{equation}
where $\Delta_{31} = \Delta_{21} \equiv \Delta_p$. This equation can be used for numerical modelling with Gaussian disorder upon standard replacement $\Sigma_{ik}\to\gamma_{ik}$. Here we have introduced the dimensionless signal parameter
\begin{equation}\label{eq:lambda_s}
\lambda_s = \frac{\Omega_s^2}{(\Delta_{41}^2 + \Sigma_{41}^2)},
\end{equation}
such that Eq. (\ref{eq:eta_full}) reduces to Eq. (\ref{eq:eta analytic}) when $\lambda_s=0$. The impact of the signal field is determined entirely by $\lambda_s$. We find that finite $\lambda_s$ result in suppression of $\eta_p$, so that the {\it probe-only} performance in Eq. (\ref{eq:eta analytic}) corresponds to the upper bound of performance of the metamaterial with both probe and signal fields being present. Such a suppression is explained by the observation that the signal field tends to close the AT transparency window (see Figure \ref{fig:AT response}b). Fortunately, reaching $\eta_p>1$ is still quite possible with small values of $\lambda_s$. For example, the dashed line in Fig. \ref{fig:disorder scaling}a  corresponds to the signal-on arrangement with $\lambda_s = 0.036$.  As Eq. (\ref{eq:lambda_s}) shows, small values of $\lambda_s$ are achieved either by reducing the strength of the signal field, or by detuning it from the $\ket{2} \to \ket{4}$ transition (since $\Delta_{41} = \Delta_{p}+\Delta_s$). Both these actions effectively reduce the detrimental effect of losses via the two-photon absorption channel TPA$_{41}$.

Our metamaterial thus implements a weak-field controllable optical nonlinearity, with the phase shift of the probe beam being controlled by the strength of the signal beam. It can also be used as an optical switch; when $\eta_p > 1$, the probe beam passing through the disordered metamaterial preserves coherence and can interfere with the reference beam, whereas $\eta_p<1$ means loss of coherence and destroyed interference. The crossover between these two regimes is reversible and is controlled by simply changing the intensity of the signal beam.

In Fig. \ref{fig:switch diagram}, we show a parameter map $(g_c,\Omega_s,\eta_{\rm max})$ for optical switching with a signal field detuned by $\Delta_s=1.0\,\gamma_{31}$. We assume a relatively large Raman decay rate $\gamma_{21}=0.5\,\gamma_{31}$, to highlight the feasibility of the optical switching scheme under realistic picocavity conditions. For a typical $S_1$ radiative lifetime $1/\gamma_{31}\sim 1\, {\rm ns}$ and inhomogeneous width $\sigma_{31}\sim 50$ meV, Fig. \ref{fig:switch diagram} shows that signal pulses with far-field intensities $I_s\sim 10 \;{\rm kW/cm}^2$ ($d_{42}\sim 1$ D), are sufficient to switch off coherent phase modulation, in an ensemble of resonant picocavities in the Autler-Townes regime ($g_c>\Sigma_{31}$). These field intensities are orders of magnitude smaller than the typical  two-photon excitation intensities used in photochemistry ($\sim 10^6$ W/cm$^2$ \cite{KOBAYASHI2018,Rumi10}) or microscopy ($I\sim10^{15}$ W/cm$^2$ \cite{SVOBODA2006,Langer2013}), which can facilitate the implementation of an optical switch based on the Autler-Townes window.

\section{Conclusions}
\label{sec:conclusions}

\notablue{We perform a proof-of-principle theoretical analysis of coherent  optical phase manipulation assisted by the electromagnetic vacuum in a dilute ensemble of disordered single-molecule plasmonic picocavities.  
We show that strong light-matter coupling with individual organic chromophores that have well-resolved vibronic progressions in the absorption and emission spectra opens an Autler-Townes transparency band \cite{Fleischhauer:2005} in the $0\rightarrow \tilde 0$ sideband of the chromophore absorption spectrum, which we propose to use for coherent manipulations of the average refractive index of the disordered medium at these frequencies, resulting in a controllable phase shift of a propagating probe wave tuned on resonance with the $0\rightarrow \tilde 0$ sideband. \notaviolet{Although the achievable refractive index variations may be small in a potential realization of the proposed scheme, plasmonic nanocavity structures can in principle be engineered to detect refractive index changes at optical frequencies of 0.1\% or less \cite{Gao2011}.}}

\notablue{Assuming a Lorentzian model for the picocavity spectrum and a displaced-oscillator model for the chromophore levels \cite{Herrera2017-PRL,Herrera2017-PRA,Herrera2016}, we obtain an analytical expression for the ratio between the dispersive and absorptive parts of the average system susceptibility at a probe frequency, which highlights the dependence of the predicted phase shift on important design parameters such as the inhomogeneous linewidth of the relevant vibronic transitions, the radiative and non-radiative molecular relaxation rates, the picocavity photon lifetime, and the average single-molecule Rabi frequency. The predicted phase signal may be challenging to detect with current plasmonic picocavities  \cite{Benz2016,Carnegie2018,Chikkaraddy:2016aa,Chikkaraddy2018}, but expected improvements in nanofabrication may enable the observation of the proposed phase control scheme. }

In our analysis we take into account realistic sources of disorder and relaxation including a distribution of Rabi couplings, energy disorder in the molecular transition frequencies, vibrational relaxation and photon losses. 
\notablue{The figure-of-merit for the proposed picocavity-induced phase shift at the probe frequency was found to be significantly more sensitive to static disorder in the molecular transition frequencies than disorder in the Rabi coupling strength. } 

The proposed phase shift of the probe field can be dynamically gated with an additional signal field at a higher frequency set to drive an excited state molecular coherence. The phase switch mechanism is interpreted as the opening of a novel two-color two-photon absorption process in the molecules, mediated and enhanced by the ensemble of picocavity vacua. Signal fields as weak as $10$ kW/cm$^2$ are estimated to be sufficient for implementing optical switching behavior in a disordered picocavity ensemble. Our work thus opens the way toward the development of few-photon nonlinear optical devices with molecular picocavity metamaterials.

\label{sec:conclusions}

\section{Acknowledgments}
F.H. is funded by ANID -- Fondecyt Regular 1181743 and Millennium Science Initiative Program ICN17\_012.

\bibliographystyle{unsrt}
\bibliography{VIT-metasurface.bib}

\begin{thebibliography}{10}

\bibitem{Benz2016}
Felix Benz, Mikolaj~K. Schmidt, Alexander Dreismann, Rohit Chikkaraddy, Yao
  Zhang, Angela Demetriadou, Cloudy Carnegie, Hamid Ohadi, Bart de~Nijs, Ruben
  Esteban, Javier Aizpurua, and Jeremy~J. Baumberg.
\newblock Single-molecule optomechanics in picocavities.
\newblock {\em Science}, 354(6313):726--729, 2016.

\bibitem{Carnegie2018}
Cloudy Carnegie, Jack Griffiths, Bart de~Nijs, Charlie Readman, Rohit
  Chikkaraddy, William~M. Deacon, Yao Zhang, Istv{\'a}n Szab{\'o}, Edina Rosta,
  Javier Aizpurua, and Jeremy~J. Baumberg.
\newblock Room-temperature optical picocavities below 1 nm3 accessing
  single-atom geometries.
\newblock {\em The Journal of Physical Chemistry Letters}, 9(24):7146--7151,
  2018.

\bibitem{Chikkaraddy:2016aa}
Rohit Chikkaraddy, Bart de~Nijs, Felix Benz, Steven~J. Barrow, Oren~A.
  Scherman, Edina Rosta, Angela Demetriadou, Peter Fox, Ortwin Hess, and
  Jeremy~J. Baumberg.
\newblock Single-molecule strong coupling at room temperature in plasmonic
  nanocavities.
\newblock {\em Nature}, 535(7610):127--130, 07 2016.

\bibitem{Chikkaraddy2018}
Rohit Chikkaraddy, V.~A. Turek, Nuttawut Kongsuwan, Felix Benz, Cloudy
  Carnegie, Tim van~de Goor, Bart de~Nijs, Angela Demetriadou, Ortwin Hess,
  Ulrich~F. Keyser, and Jeremy~J. Baumberg.
\newblock Mapping nanoscale hotspots with single-molecule emitters assembled
  into plasmonic nanocavities using dna origami.
\newblock {\em Nano Letters}, 18(1):405--411, 2018.
\newblock PMID: 29166033.

\bibitem{Ebbesen2016}
Thomas~W Ebbesen.
\newblock {Hybrid Light-Matter States in a Molecular and Material Science
  Perspective}.
\newblock {\em Accounts of Chemical Research}, 49:2403--2412, 2016.

\bibitem{Herrera2020perspective}
Felipe Herrera and Jeffrey Owrutsky.
\newblock Molecular polaritons for controlling chemistry with quantum optics.
\newblock {\em The Journal of Chemical Physics}, 152(10):100902, 2020.

\bibitem{Zhang2021}
Yuan Zhang, Ruben Esteban, Roberto~A. Boto, Mattin Urbieta, Xabier Arrieta,
  ChongXin Shan, Shuzhou Li, Jeremy~J. Baumberg, and Javier Aizpurua.
\newblock Addressing molecular optomechanical effects in nanocavity-enhanced
  raman scattering beyond the single plasmonic mode.
\newblock {\em Nanoscale}, 13:1938--1954, 2021.

\bibitem{Delga2014}
A~Delga, J~Feist, J~Bravo-Abad, and F~J Garcia-Vidal.
\newblock Theory of strong coupling between quantum emitters and localized
  surface plasmons.
\newblock {\em Journal of Optics}, 16(11):114018, 2014.

\bibitem{Neuman2018}
Tom\'{a}\v{s} Neuman and Javier Aizpurua.
\newblock Origin of the asymmetric light emission from molecular
  exciton-polaritons.
\newblock {\em Optica}, 5(10):1247--1255, Oct 2018.

\bibitem{Neuman2020}
Tom{\'a}{\v s} Neuman, Javier Aizpurua, and Ruben Esteban.
\newblock Quantum theory of surface-enhanced resonant raman scattering (serrs)
  of molecules in strongly coupled plasmon--exciton systems.
\newblock {\em Nanophotonics}, 9(2):295--308, 2020.

\bibitem{Feist2021}
Johannes Feist, Antonio~I. Fern{\'a}ndez-Dom{\'\i}nguez, and Francisco~J.
  Garc{\'\i}a-Vidal.
\newblock Macroscopic qed for quantum nanophotonics: emitter-centered modes as
  a minimal basis for multiemitter problems.
\newblock {\em Nanophotonics}, 10(1):477--489, 2021.

\bibitem{Schmid2013}
Thomas Schmid, Lothar Opilik, Carolin Blum, and Renato Zenobi.
\newblock {Nanoscale chemical imaging using tip-enhanced raman spectroscopy: A
  critical review}.
\newblock {\em Angewandte Chemie - International Edition}, 52(23):5940--5954,
  2013.

\bibitem{Tame2013}
M.~S. Tame, K.~R. McEnery, {\c S}.~K. {\"{O}}zdemir, J.~Lee, S.~a. Maier, and
  M.~S. Kim.
\newblock {Quantum plasmonics}.
\newblock {\em Nature Physics}, 9(6):329--340, jun 2013.

\bibitem{Agranovich2005}
V.M. Agranovich and G.C.~La Rocca.
\newblock Electronic excitations in organic microcavities with strong
  light--matter coupling.
\newblock {\em Solid State Communications}, 135(9--10):544 -- 553, 2005.
\newblock Fundamental Optical and Quantum Effects in Condensed Matter.

\bibitem{Tischler2007}
J.~R. Tischler~{\it et al.}
\newblock Solid state cavity qed: Strong coupling in organic thin films.
\newblock {\em Organic Electronics}, 8(2--3):94 -- 113, 2007.

\bibitem{kena-cohen2008}
S.~K\'{e}na-Cohen, M.~Davan\c{c}o, and S.~R. Forrest.
\newblock Strong exciton-photon coupling in an organic single crystal
  microcavity.
\newblock {\em Phys. Rev. Lett.}, 101:116401, Sep 2008.

\bibitem{Herrera2017-PRA}
Felipe Herrera and Frank~C. Spano.
\newblock Absorption and photoluminescence in organic cavity qed.
\newblock {\em Phys. Rev. A}, 95:053867, May 2017.

\bibitem{Herrera2017-PRL}
Felipe Herrera and Frank~C. Spano.
\newblock Dark vibronic polaritons and the spectroscopy of organic
  microcavities.
\newblock {\em Phys. Rev. Lett.}, 118:223601, May 2017.

\bibitem{Mazzeo2014}
M.~Mazzeo~{\it et al.}
\newblock Ultrastrong light-matter coupling in electrically doped microcavity
  organic light emitting diodes.
\newblock {\em Appl. Phys. Lett.}, 104(23), 2014.

\bibitem{Gambino2015}
S.~Gambino~{\it et al.}
\newblock Ultrastrong light-matter coupling in electroluminescent organic
  microcavities.
\newblock {\em Applied Materials Today}, 1(1):33 -- 36, 2015.

\bibitem{Schartz2011}
T.~Schwartz, J.~A. Hutchison, C.~Genet, and T.~W. Ebbesen.
\newblock Reversible switching of ultrastrong light-molecule coupling.
\newblock {\em Phys. Rev. Lett.}, 106:196405, 2011.

\bibitem{Kena-Cohen2013}
St{\'e}phane K{\'e}na-Cohen, Stefan~A. Maier, and Donal D.~C. Bradley.
\newblock Ultrastrongly coupled exciton--polaritons in metal-clad organic
  semiconductor microcavities.
\newblock {\em Advanced Optical Materials}, 1(11):827--833, 2013.

\bibitem{Hobson2002}
P.~A. Hobson, W.~L. Barnes, D.~G. Lidzey, G.~A. Gehring, D.~M. Whittaker, M.~S.
  Skolnick, and S.~Walker.
\newblock Strong exciton--photon coupling in a low-{Q} all-metal mirror
  microcavity.
\newblock {\em Applied Physics Letters}, 81(19):3519--3521, 2002.

\bibitem{Feist2015}
J.~Feist and F.~J. Garcia-Vidal.
\newblock Extraordinary exciton conductance induced by strong coupling.
\newblock {\em Phys. Rev. Lett.}, 114:196402, 2015.

\bibitem{Schachenmayer2015}
J.~Schachenmayer, C.~Genes, E.~Tignone, and G.~Pupillo.
\newblock Cavity-enhanced transport of excitons.
\newblock {\em Phys. Rev. Lett.}, 114:196403, 2015.

\bibitem{yuen2016}
Joel Yuen-Zhou, Semion~K Saikin, Tony Zhu, Mehmet~C Onbasli, Caroline~A Ross,
  Vladimir Bulovic, and Marc~A Baldo.
\newblock Plexciton dirac points and topological modes.
\newblock {\em Nature Communications}, 7:11783, 2016.

\bibitem{Pupillo2017}
David Hagenm\"uller, Johannes Schachenmayer, Stefan Sch\"utz, Claudiu Genes,
  and Guido Pupillo.
\newblock Cavity-enhanced transport of charge.
\newblock {\em Phys. Rev. Lett.}, 119:223601, Nov 2017.

\bibitem{Hutchison2012}
James~A Hutchison, Tal Schwartz, Cyriaque Genet, E.~Devaux, and T.~W. Ebbesen.
\newblock {Modifying Chemical Landscapes by Coupling to Vacuum Fields}.
\newblock {\em Angew. Chem. Int. Ed.}, 51(7):1592--1596, 2012.

\bibitem{Herrera2016}
Felipe Herrera and Frank~C. Spano.
\newblock Cavity-controlled chemistry in molecular ensembles.
\newblock {\em Phys. Rev. Lett.}, 116:238301, Jun 2016.

\bibitem{Herrera2014}
F.~Herrera~{\it et al.}
\newblock Quantum nonlinear optics with polar {J}-aggregates in microcavities.
\newblock {\em J. Phys. Chem. Lett.}, 5(21):3708--3715, 2014.

\bibitem{Barachati2018}
F{\'a}bio Barachati, Janos Simon, Yulia~A. Getmanenko, Stephen Barlow, Seth~R.
  Marder, and St{\'e}phane K{\'e}na-Cohen.
\newblock Tunable third-harmonic generation from polaritons in the ultrastrong
  coupling regime.
\newblock {\em ACS Photonics}, 5(1):119--125, 2018.

\bibitem{Daskalakis2017}
Konstantinos~S Daskalakis, Stefan~A Maier, and St{\'e}phane K{\'e}na-Cohen.
\newblock Polariton condensation in organic semiconductors.
\newblock In {\em Quantum Plasmonics}, pages 151--163. Springer International
  Publishing, 2017.

\bibitem{Lerario2017}
Giovanni Lerario, Antonio Fieramosca, F{\'a}bio Barachati, Dario Ballarini,
  Konstantinos~S Daskalakis, Lorenzo Dominici, Milena De~Giorgi, Stefan~A
  Maier, Giuseppe Gigli, and St{\'e}phane K{\'e}na-Cohen.
\newblock Room-temperature superfluidity in a polariton condensate.
\newblock {\em Nature Physics}, (13):837–841, 2017.

\bibitem{Zhu:2015}
B~Zhu, J~Schachenmayer, M~Xu, F~Herrera, J~G Restrepo, M~J Holland, and A~M
  Rey.
\newblock Synchronization of interacting quantum dipoles.
\newblock {\em New Journal of Physics}, 17(8):083063, 2015.

\bibitem{Baranov2017}
Denis~G. Baranov, Martin Wers{\"a}ll, Jorge Cuadra, Tomasz~J. Antosiewicz, and
  Timur Shegai.
\newblock Novel nanostructures and materials for strong light--matter
  interactions.
\newblock {\em ACS Photonics}, 5:24--42, 2018.

\bibitem{Muller2018}
Eric~A. Muller, Benjamin Pollard, Hans~A. Bechtel, Ronen Adato, Dordaneh
  Etezadi, Hatice Altug, and Markus~B. Raschke.
\newblock Nanoimaging and control of molecular vibrations through
  electromagnetically induced scattering reaching the strong coupling regime.
\newblock {\em ACS Photonics}, 5(9):3594--3600, 09 2018.

\bibitem{Felicetti2020}
Simone Felicetti, Jacopo Fregoni, Thomas Schnappinger, Sebastian Reiter, Regina
  de~Vivie-Riedle, and Johannes Feist.
\newblock Photoprotecting uracil by coupling with lossy nanocavities.
\newblock {\em The Journal of Physical Chemistry Letters}, 11(20):8810--8818,
  2020.

\bibitem{Fregoni2018}
Jacopo Fregoni, Giovanni Granucci, Emanuele Coccia, Maurizio Persico, and
  Stefano Corni.
\newblock Manipulating azobenzene photoisomerization through strong
  light--molecule coupling.
\newblock {\em Nature communications}, 9(1):1--9, 2018.

\bibitem{Behr2008}
Nicolas Behr and Markus~B. Raschke.
\newblock Optical antenna properties of scanning probe tips.
\newblock {\em The Journal of Physical Chemistry C}, 112(10):3766--3773, 03
  2008.

\bibitem{May2020}
Molly~A. May, David Fialkow, Tong Wu, Kyoung-Duck Park, Haixu Leng, Jaron~A.
  Kropp, Theodosia Gougousi, Philippe Lalanne, Matthew Pelton, and Markus~B.
  Raschke.
\newblock Nano-cavity qed with tunable nano-tip interaction.
\newblock {\em Advanced Quantum Technologies}, 3(2):1900087, 2020.

\bibitem{Metzger2019}
Bernd Metzger, Eric Muller, Jun Nishida, Benjamin Pollard, Mario Hentschel, and
  Markus~B. Raschke.
\newblock Purcell-enhanced spontaneous emission of molecular vibrations.
\newblock {\em Phys. Rev. Lett.}, 123:153001, Oct 2019.

\bibitem{Litinskaya2019}
Marina Litinskaya and Felipe Herrera.
\newblock Vacuum-enhanced optical nonlinearities with disordered molecular
  photoswitches.
\newblock {\em Phys. Rev. B}, 99:041107, Jan 2019.

\bibitem{Spano2011}
Frank~C Spano and Hajime Yamagata.
\newblock {Vibronic coupling in J-aggregates and beyond: a direct means of
  determining the exciton coherence length from the photoluminescence
  spectrum.}
\newblock {\em The journal of physical chemistry. B}, 115(18):5133--43, may
  2011.

\bibitem{Bhawalkar1996}
J~D Bhawalkar, G~S He, and P~N Prasad.
\newblock Nonlinear multiphoton processes in organic and polymeric materials.
\newblock {\em Reports on Progress in Physics}, 59(9):1041, 1996.

\bibitem{Lassiter2013}
J.~Britt Lassiter, Felicia McGuire, Jack~J. Mock, Cristian Cirac{\`\i}, Ryan~T.
  Hill, Benjamin~J. Wiley, Ashutosh Chilkoti, and David~R. Smith.
\newblock Plasmonic waveguide modes of film-coupled metallic nanocubes.
\newblock {\em Nano Letters}, 13(12):5866--5872, 12 2013.

\bibitem{Esashika:19}
Keiko Esashika, Ryo Ishii, Shunya Tokihiro, and Toshiharu Saiki.
\newblock Simple and rapid method for homogeneous dimer formation of gold
  nanoparticles in a bulk suspension based on van der waals interactions
  between alkyl chains.
\newblock {\em Opt. Mater. Express}, 9(4):1667--1677, Apr 2019.

\bibitem{Jais2011}
Pablo~M. Jais, Daniel~B. Murray, Roberto Merlin, and Andrea~V. Bragas.
\newblock Metal nanoparticle ensembles: Tunable laser pulses distinguish
  monomer from dimer vibrations.
\newblock {\em Nano Letters}, 11(9):3685--3689, 09 2011.

\bibitem{Babonneau2018}
D~Babonneau, D~K Diop, L~Simonot, B~Lamongie, N~Blanc, N~Boudet, F~Vocanson,
  and N~Destouches.
\newblock Real-time investigations of structural and optical changes in
  photochromic ag/{TiO}2 nanocomposite thin films under laser irradiation.
\newblock {\em Nano Futures}, 2(1):015002, feb 2018.

\bibitem{Spano2010}
Frank~C Spano.
\newblock The spectral signatures of frenkel polarons in h- and j-aggregates.
\newblock {\em Acc. Chem. Res.}, 43(3):429--439, 2010.

\bibitem{Sihvola-book}
Ari Sihvola.
\newblock {\em Electromagnetic Mixing Formulas and Applications}.
\newblock Electromagnetic Waves. Institution of Engineering and Technology,
  1999.

\bibitem{ASPNES2011}
D.E. Aspnes.
\newblock Plasmonics and effective-medium theories.
\newblock {\em Thin Solid Films}, 519(9):2571--2574, 2011.
\newblock 5th International Conference on Spectroscopic Ellipsometry (ICSE-V).

\bibitem{Markel2016}
Vadim~A. Markel.
\newblock Introduction to the maxwell garnett approximation: tutorial.
\newblock {\em J. Opt. Soc. Am. A}, 33(7):1244--1256, Jul 2016.

\bibitem{Czajkowski2018}
Krzysztof~M. Czajkowski, Dominika {\'S}witlik, Christoph Langhammer, and
  Tomasz~J. Antosiewicz.
\newblock Effective optical properties of inhomogeneously distributed
  nanoobjects in strong field gradients of nanoplasmonic sensors.
\newblock {\em Plasmonics}, 13(6):2423--2434, 2018.

\bibitem{Lifshitz1}
I.M. Lifshitz and L.N. Rozenzweig.
\newblock On elastic properties of polycrystals.
\newblock {\em Zh. Exp. Teor. Fiz.}, (16):967.

\bibitem{Lifshitz2}
I.M. Lifshitz, M.I. Kaganov, and V.M. Tzukernic.
\newblock Propagation of electromagnetic vibrations in non-uniform anisotropic
  media; also available in: Selected works of i.m. lifshitz, nauka, moscow,
  1987, p. 337.
\newblock {\em Uch. Zap. KhGU}, (2):41--54.

\bibitem{Kaganova2003}
Inna~M. Kaganova.
\newblock On calculation of effective conductivity of inhomogeneous metals.
\newblock {\em Physics Letters A}, 312(1):108--118, 2003.

\bibitem{Kaganova1995}
I.~M. Kaganova.
\newblock Theory of surface polaritons in polycrystals.
\newblock {\em Phys. Rev. B}, 51:5333--5344, Feb 1995.

\bibitem{LITINSKAIA2000}
M.L. Litinskaia and I.M. Kaganova.
\newblock Motional narrowing in a microcavity: contribution to the lower
  polariton linewidth.
\newblock {\em Physics Letters A}, 275(4):292--298, 2000.

\bibitem{Fojt2021}
Jakub Fojt, Tuomas~P. Rossi, Tomasz~J. Antosiewicz, Mikael Kuisma, and Paul
  Erhart.
\newblock Dipolar coupling of nanoparticle-molecule assemblies: An efficient
  approach for studying strong coupling.
\newblock {\em The Journal of Chemical Physics}, 154(9):094109, 2021.

\bibitem{Xu2019-sensing}
Yi~Xu, Ping Bai, Xiaodong Zhou, Yuriy Akimov, Ching~Eng Png, Lay-Kee Ang,
  Wolfgang Knoll, and Lin Wu.
\newblock Optical refractive index sensors with plasmonic and photonic
  structures: Promising and inconvenient truth.
\newblock {\em Advanced Optical Materials}, 7(9):1801433, 2019.

\bibitem{Papaioannou2016}
Maria Papaioannou, Eric Plum, Jo{\~a}o Valente, Edward~TF Rogers, and Nikolay~I
  Zheludev.
\newblock Two-dimensional control of light with light on metasurfaces.
\newblock {\em Light: Science \& Applications}, 5(4):e16070--e16070, 2016.

\bibitem{Kessing2021}
R.~Kevin Kessing, Pei-Yun Yang, Salvatore~R. Manmana, and Jianshu Cao.
\newblock Long-range non-equilibrium coherent tunneling induced by fractional
  vibronic resonances, 2021.

\bibitem{Phillips2011}
Nathaniel~B. Phillips, Alexey~V. Gorshkov, and Irina Novikova.
\newblock Light storage in an optically thick atomic ensemble under conditions
  of electromagnetically induced transparency and four-wave mixing.
\newblock {\em Phys. Rev. A}, 83:063823, Jun 2011.

\bibitem{Engelhardt2021prl}
Georg Engelhardt and Jianshu Cao.
\newblock Dynamical symmetries and symmetry-protected selection rules in
  periodically driven quantum systems.
\newblock {\em Phys. Rev. Lett.}, 126:090601, Mar 2021.

\bibitem{Fleischhauer:2005}
M.~Fleischhauer, A.~Imamoglu, and J.~P. Marangos.
\newblock {Electromagnetically Induced Transparency: Optics in Coherent Media}.
\newblock {\em Rev. Mod. Phys.}, 77:633--673, Jul 2005.

\bibitem{Casellas2016}
Josep Casellas, Michael~J. Bearpark, and Mar Reguero.
\newblock Excited-state decay in the photoisomerisation of azobenzene: A new
  balance between mechanisms.
\newblock {\em ChemPhysChem}, 17(19):3068--3079, 2016.

\bibitem{Maier2006}
Stefan~A Maier.
\newblock Plasmonic field enhancement and sers in the effective mode volume
  picture.
\newblock {\em Optics Express}, 14(5):1957--1964, 2006.

\bibitem{Pollard2014}
Benjamin Pollard, Eric~A. Muller, Karsten Hinrichs, and Markus~B. Raschke.
\newblock Vibrational nano-spectroscopic imaging correlating structure with
  intermolecular coupling and dynamics.
\newblock {\em Nature Communications}, 5(1):3587, 2014.

\bibitem{Vlaming2009}
S.~M. Vlaming, V.~A. Malyshev, and J.~Knoester.
\newblock Localization properties of one-dimensional frenkel excitons: Gaussian
  versus lorentzian diagonal disorder.
\newblock {\em Phys. Rev. B}, 79:205121, May 2009.

\bibitem{Engelhardt2021}
Georg Engelhardt and Jianshu Cao.
\newblock Unusual dynamical properties of disordered polaritons in
  micocavities, 2021.

\bibitem{Wang2017}
Daqing Wang, Hrishikesh Kelkar, Diego Martin-Cano, Tobias Utikal, Stephan
  G{\"{o}}tzinger, and Vahid Sandoghdar.
\newblock {Coherent Coupling of a Single Molecule to a Scanning Fabry-Perot
  Microcavity}.
\newblock {\em Physical Review X}, 7(2):021014, apr 2017.

\bibitem{KOBAYASHI2018}
Yoichi Kobayashi, Katsuya Mutoh, and Jiro Abe.
\newblock Stepwise two-photon absorption processes utilizing photochromic
  reactions.
\newblock {\em Journal of Photochemistry and Photobiology C: Photochemistry
  Reviews}, 34:2--28, 2018.

\bibitem{Rumi10}
Mariacristina Rumi and Joseph~W. Perry.
\newblock Two-photon absorption: an overview of measurements and principles.
\newblock {\em Adv. Opt. Photon.}, 2(4):451--518, Dec 2010.

\bibitem{SVOBODA2006}
Karel Svoboda and Ryohei Yasuda.
\newblock Principles of two-photon excitation microscopy and its applications
  to neuroscience.
\newblock {\em Neuron}, 50(6):823--839, 2006.

\bibitem{Langer2013}
Gregor Langer, Klaus-Dieter Bouchal, Hubert Gr\"{u}n, Peter Burgholzer, and
  Thomas Berer.
\newblock Two-photon absorption-induced photoacoustic imaging of rhodamine b
  dyed polyethylene spheres using a femtosecond laser.
\newblock {\em Opt. Express}, 21(19):22410--22422, Sep 2013.

\bibitem{Gao2011}
Yongkang Gao, Qiaoqiang Gan, Zheming Xin, Xuanhong Cheng, and Filbert~J.
  Bartoli.
\newblock Plasmonic mach--zehnder interferometer for ultrasensitive on-chip
  biosensing.
\newblock {\em ACS Nano}, 5(12):9836--9844, 2011.
\newblock PMID: 22067195.

\end{thebibliography}

\begin{widetext}
\appendix

\section{Derivation of the Effective Index at the Probe Frequency}
\label{app:effective index}

Here we derive Eq. (\ref{eq:n full}) from the main text, following the procedure created in Refs.~\cite{Lifshitz1,Lifshitz2} for various kinds of random media that allow for perturbative approach. Our starting point is Eq.(\ref{eq:local wave eq1}). Its $n$-th discrete frequency component oscillating at the frequency $\omega_n$ satisfies the wave equation:
\begin{eqnarray}\label{eq:wave eq}
\nabla^2\E_n(\r) +\frac{\epsilon_{\rm d}\omega_n^2}{c^2}\,\mathbf{E}_n(\r) &=& -\frac{\omega_n^2}{\epsilon_0c^2}\mathbf{P}_{n}(\r),
\end{eqnarray}
where we used $\D_{\rm d}(\r)=\epsilon_0\epsilon_{\rm d}\E(\r)$, and $\P_n(\r)$ is the component of the polarization density at $\omega_n$, caused by inclusions (cavities with embedded point dipoles). We assume that at the probe frequency ($\omega_n=\omega_p$) a linear relationship between the polarization and the (weak) probe field holds, so that 
\begin{equation}\label{eq:P probe}
\P_p(\r) =\epsilon_0\chi_L(\omega_p,\r)\cdot \E_p(\r),
\end{equation}
where the susceptibility $\chi=\chi(\omega,\r)$ captures the fact that the polarization is created locally, and vanishes between the picocavities.

Following the effective medium approach from Refs. \cite{Lifshitz1,Lifshitz2,Kaganova2003,Kaganova1995,LITINSKAIA2000}, we write the electric field and the susceptibility as
\begin{eqnarray}\label{eq:Ep}
\E_p(\r)&=&\langle \E_p(\r)\rangle + \delta\E_p(\r),\label{eq:local Ep}\\
\chi(\r) & =& \langle \chi\rangle + \delta \chi(\r),\label{eq:local chi}
\end{eqnarray}
where  $\langle \E_p(\r)\rangle$ is the average probe field  that propagates according to the effective index $n(\omega_p)$, and $\delta\E_p(\r)$ is a position-dependent fluctuation of the electric field caused by the presence of picocavities; $\langle \chi\rangle$ is the uniform effective susceptibility of the medium (the effective medium correction), and $ \delta \chi(\r)$ is the local fluctuation of the response. Our mean values are chosen in such a way that, by construction,  
\begin{equation}\label{eq:zerolevel}
\langle \delta \E_p(r)\rangle =\langle \delta \chi(\r)\rangle=0.
\end{equation}

Inserting Eqs. (\ref{eq:P probe}), (\ref{eq:local Ep}), and (\ref{eq:local chi}) into Eq. (\ref{eq:wave eq}), and averaging the result, we obtain an equation for the averages, which reads:
\begin{equation}\label{eq:averages}
\nabla^2\langle \E(\r)\rangle + \frac{\omega_p^2}{c^2}\left[\epsilon_{\rm d}+\langle \chi\rangle \right]\langle \E(\r)\rangle= -\frac{\omega_p^2}{\epsilon_0 c^2}\langle \delta \chi(\r)\,\delta\E(\r)\rangle,
\end{equation}
where we used (\ref{eq:zerolevel}) to eliminate the terms proportional to $\delta\chi$ and $\delta \E$ times a position-independent factor. We then subtract (\ref{eq:averages}) from our starting-point equation [Eq.(\ref{eq:wave eq}) with (\ref{eq:P probe}), (\ref{eq:local Ep}), and (\ref{eq:local chi}) plugged in]. This leaves us with an equation for the fluctuations: 
\begin{equation}\label{eq:fluctuations}
\nabla^2\,\delta\E(\r)+ \frac{\omega_p^2}{c^2}\left[\epsilon_{\rm d}+\langle \chi\rangle \right]\delta \E(\r)= -\frac{\omega_p^2}{\epsilon_0 c^2}\,\delta \chi(\r)\,\langle \E(\r)\rangle- \frac{\omega_p^2}{\epsilon_0 c^2}\left[\delta\chi(\r)\delta\E(\r)-\langle \delta\chi(\r)\delta\E(\r)\rangle\right].
\end{equation}

For now we will assume that the role of the fluctuations is relatively small; we will quantify this assumption below. This allows us to neglect the second-order term $\propto \left[\delta\chi(\r)\delta\E(\r)-\langle \delta\chi(\r)\delta\E(\r)\rangle\right]$ in the last equation. Our position-dependent first-order equations describing fields in the disordered medium thus become:

\begin{eqnarray} 
\nabla^2\langle \E(\r)\rangle + \frac{\omega_p^2}{c^2} \left[\epsilon_{\rm d}+\langle \chi\rangle \right]\langle \E(\r)\rangle &=& -\frac{\omega_p^2}{c^2}\langle \delta \chi(\r)\,\delta\E(\r)\rangle,\label{eq:Lifsh1a}\\
\nabla^2\,\delta\E(\r) + \frac{\omega_p^2}{c^2}\left[\epsilon_{\rm d}+\langle \chi\rangle \right]\delta \E(\r) &=& -\frac{\omega_p^2}{c^2}\,\delta \chi(\r)\,\langle \E(\r)\rangle.\label{eq:Lifsh1b}
\end{eqnarray}

We keep the second-order term in the equation (\ref{eq:Lifsh1a}) for the averages, since there it is the lowest-order disorder-dependent contribution. 

The next step is to Fourier-transform these equations. This gives:

\begin{eqnarray}
\left( \frac{\omega_p^2}{c^2}[\epsilon_d + \langle \chi\rangle]  - k^2 \right) \langle \E({\bf k})\rangle &=& -\frac{\omega_p^2}{\epsilon_0 c^2} \int d{\bf q} \langle \delta\chi({\bf k}-{\bf q}) \delta\E({\bf q})\rangle;\label{eq:Lifsh2a}\\
\left( \frac{\omega_p^2}{c^2}[\epsilon_d + \langle \chi\rangle] - k^2 \right) \delta\E({\bf k}) &=& -\frac{\omega_p^2}{\epsilon_0 c^2} \int d{\bf q}' \delta\chi({\bf k}-{\bf q}') \langle \E({\bf q}')\rangle.\label{eq:Lifsh2b}
\end{eqnarray} 

We solve Eq.(\ref{eq:Lifsh2b}) for $\delta\E({\bf k})$, and plug it into the integrand in Eq.(\ref{eq:Lifsh2a}). We get:

\begin{equation}\label{eq:almost_there}
    \left( \frac{\omega_p^2}{c^2}[\epsilon_d + \langle \chi\rangle]  - k^2 \right) \langle \E({\bf k})\rangle = \left(\frac{\omega_p^2}{\epsilon_0 c^2}\right)^2 \int\int d{\bf q}\ d{\bf q}' \frac{\langle \E({\bf q}')\rangle}{\left(\frac{\omega_p^2}{c^2}[\epsilon_d + \langle \chi\rangle] - q^2\right) }\langle \delta\chi({\bf k}-{\bf q})\delta\chi({\bf q}-{\bf q}') \rangle.
\end{equation}
  
This integral equation for $\langle \E(\bf k) \rangle$ shows that the field average depends on two deterministic components: the propagator of electromagnetic waves in the effective medium, $G(q) = \left(\frac{\omega_p^2}{c^2}[\epsilon_d + \langle \chi\rangle] - q^2\right)^{-1}$, and the correlator $\langle \delta\chi({\bf k}-{\bf q})\delta\chi({\bf k}-{\bf q}') \rangle$. We can simplify it further by examining the properties of this correlator. Let us denote as 
\begin{equation}
    K(|\r_1-\r_2|) = \langle \delta\chi(\r_1) \delta\chi(\r_2) \rangle
\end{equation}
the spatial correlator of the susceptibility fluctuations. This form is very general and only assumes that the medium is homogeneous and isotropic on average (and hence the correlator depends only on $|\r_1-\r_2|$). We can now write: 

\begin{eqnarray}
\langle \delta\chi({\bf k}-{\bf q})\delta\chi({\bf q}-{\bf q}') \rangle &=& \frac{1}{(2\pi)^6} \int\int d\r_1 d\r_2 \ e^{-i({\bf k}-{\bf q})\r_1}\ e^{-i({\bf q}-{\bf q}')\r_2}K(|\r_1-\r_2|)\\
&=& \frac{1}{(2\pi)^6} \int d\r_2 \ e^{-i({\bf k}-{\bf q}')\r_2}\int d(\r_1-\r_2) \ e^{-i({\bf k}-{\bf q})(\r_1-\r_2)}K(|\r_1-\r_2|)\\
&=& \delta({\bf k}-{\bf q}') K(|{\bf k}-{\bf q}|) \label{eq:delta-function}
\end{eqnarray}
with 
\begin{equation}
    K(q) = \frac{1}{(2\pi)^3} \int d\r~e^{-i{\bf q}\r} K(r)
\end{equation}
being the Fourier transform of the spatial correlator. 

We now plug Eq.(\ref{eq:delta-function}) into Eq.(\ref{eq:almost_there}) and carry out integration over ${\bf q}'$. Thanks to the delta-function, $\langle \E({\bf q}')\rangle$ transforms into $\langle \E({\bf k})\rangle$ and cancels out in the both sides of this equation. This happens since the system is homogeneous on average. Therefore, instead of an integral equation involving Fourier components of the field amplitude, we are left simply with the following modified dispersion equation of electromagnetic waves in our disordered medium:
\begin{equation}\label{eq:correlation_final}
     \left( \frac{\omega_p^2}{c^2}[\epsilon_d + \langle \chi\rangle]  - k^2 \right) = \left(\frac{\omega_p^2}{\epsilon_0 c^2}\right)^2 \int d{\bf q}~G(q)K(|{\bf k}-{\bf q}|) \equiv \frac{\omega_p^2}{c^2}\Delta\epsilon_p,
\end{equation}
where $\Delta\epsilon_p = \omega_p^2/(c^2\epsilon_0^2)\int d{\bf q}~G(q)K(|{\bf k}-{\bf q}|)$ is a deterministic, coordinate-independent correction to the averaged susceptibility $\langle \chi \rangle$. 

The discussion in the text is done in the effective medium approximation, under the assumption that $\Delta\epsilon_p$ is negligibly small. Eq.(\ref{eq:correlation_final}) allows one to verify this assumption for a given set of parameters relevant to their system, and to quantify the error associated with this assumption. The full procedure requires (i) introducing a sample-specific form of the correlator $K(r)$, (ii) as well as an infinitesimally small damping, in order to remove the pole due to the Green's function from the integration axis, and (iii) carrying the integration out explicitly. Below we will only show that having this correction small is equivalent to the requirement of having a wavelength, $\lambda$, large in comparison with the average scale $r_0$, at which the fluctuations happen. 

Consider, for example, Gaussian spatial correlations: $K(r) = K_0~e^{-r^2/(2r_0^2)}$, where the constant $K_0$ expresses the magnitude of the correlator, and $r_0$ is the scale of the order of the inclusion size; by choosing it this way we basically say that the correlations between the electric field and induced susceptibility vanish outside of each picocavity. Then the magnitude of its Fourier transform $K(q) = K_0~e^{-q^2 r_0^2/2}$ is indeed determined by the ratio $(r_0/\lambda)^2$ in the exponent, which vanishes as long as for the typical wavelength $\lambda = 2\pi/q \gg r_0$. Based on the analysis of Ref.\cite{LITINSKAIA2000}, that shows that the results of this approach are not sensitive to the specific choice of the correlator, we can argue that it is a general, correlator-independent property of our metamaterial.

\section{Derivation of the Effective Probe Susceptibility}
\label{app:chi}

Here we outline the steps in the derivation of Eq. (\ref{eq:chi generic}) in the main text, following closely the method used in Ref. \cite{Litinskaya2019}. In order to describe light-matter interaction in an effective four-level vibronic state manifold, the molecular system Hamiltonian $\hat{\mathcal{H}}$ from Eq. (\ref{eq:Hamiltonian}) is used in a Lindblad quantum master equation [Eq. (\ref{eq:Lindblad})], which reads
\begin{equation}\label{app:qme}
\frac{d}{dt}\hat \rho = -i[\hat{\mathcal{H}},\hat\rho] +{\mathcal L}_\kappa[\hat \rho]+{\mathcal L}_{\gamma_v}[\hat\rho]+{\mathcal L}_{\gamma_{v'}}[\hat\rho]+{\mathcal L}_{\gamma_e}[\hat\rho]
\end{equation}
with the Lindblad operators ${\mathcal L}$ having the decay timescales listed in Table \ref{tab:lindblad}.  ${\mathcal L}_\kappa[\hat \rho]$ describes photon decay within the picocavity, with $\kappa \sim 10-100~{\rm fs}^{-1}$ being the fastest decay timescale in the problem. ${\mathcal L}_{\gamma_v}[\hat \rho]$ and ${\mathcal L}_{\gamma_{v'}}[\hat \rho]$ correspond to  intramolecular vibration-assisted relaxation within the potentials $S_0$ to $S_1$, respectively, for decay times in the picosecond regime. The term ${\mathcal L}_{\gamma_e}[\hat\rho]$ describes the decay of the lowest electronic singlet excitation at rate $\gamma_e$. The corresponding dissipators are given by:
\begin{equation}\label{app:S decay}
\begin{array}{c}
{\mathcal L}_\kappa[\hat \rho] =({\kappa}/{2})\left(2\hat a\hat \rho\hat a^\dagger - \hat a^\dagger \hat a \hat \rho -\rho_s\hat a^\dagger \hat a\right),\\
{\mathcal L}_{\gamma_v}[\hat \rho]=({\gamma_{v}}/{2})\left( 2\ket{1}\bra{2}\hat \rho\ket{2}\bra{1} - \ket{2}\bra{2}\hat \rho - \hat \rho  \ket{2}\bra{2}\right),\\
{\mathcal L}_{\gamma_{v'}}[\hat \rho]=({\gamma_{v'}}/{2})\left( 2\ket{3}\bra{4}\hat \rho\ket{4}\bra{3} - \ket{4}\bra{4}\hat \rho - \hat \rho  \ket{4}\bra{4}\right),\\
{\mathcal L}_{\gamma_e}[\hat \rho] = ({\gamma_{e}}/{2})\left( 2\ket{1}\bra{3}\hat \rho\ket{3}\bra{1} - \ket{3}\bra{3} \hat \rho-\hat \rho \ket{3} \bra{3} + 2\ket{2}\bra{3}\hat \rho\ket{3}\bra{2} - \ket{3}\bra{3} \hat \rho - \hat \rho\ket{3}\bra{3}\right).
\end{array}
\end{equation}
In order to obtain Eq (\ref{eq:chi generic}), we focus on matrix elements $\hat \rho$ that explicitly accounts for the presence or absence of the cavity photon and introduce the notation $\rho_{ij}^{mn}(t) = \bra{i; m_c}\hat \rho(t)\ket{j; n_c}$,  where $\ket{i}$ and $\ket{j}$ represent molecular states ($i,j=1, 2,3, 4$), and $\ket{m_c}$ and $\ket{n_c}$ represent cavity Fock states with photon numbers $m_c$ and $n_c$, respectively. In order to remove fast oscillations from the equations of motion, we define slowly-varying amplitudes $\sigma_{ij}^{mn}$ for selected elements of reduced density matrix as follows: $\sigma_{13}^{00}= \rho_{13}^{00} \,{\rm e}^{-i\omega_p t}$, $\sigma_{12}^{01}={\rm e}^{-i\omega_{p}t}\rho_{12}^{01}$, $\sigma_{32}^{01}=\rho_{32}^{01}$, $\sigma_{14}^{01}={\rm e}^{-i(\omega_{p}+\omega_s)t}\rho_{14}^{01}$, $\sigma_{34}^{01}={\rm e}^{-i\omega_{s}t}\rho_{34}^{01}$, and $\sigma_{24}^{11}={\rm e}^{-i\omega_{s}t}\rho_{24}^{11}$. In terms of these slowly-varying amplitudes, we obtain from Eq.(\ref{app:qme}) the following equations of motion for the coherences:

\begin{equation}
\begin{array}{rcl}
\dot{\sigma}_{13}^{00} &=& i(\omega_{31}-\omega_p)\,\sigma_{13}^{00} -\gamma_{31}\,\sigma_{13}^{00}-i\Omega_{\rm p}(\sigma_{33}^{00}-\sigma_{11}^{00})+ig_c\sigma_{12}^{01}\;\;\;  \\

&&\\

\dot{\sigma}_{12}^{01} &=& i(\omega_{21}+\omega_c-\omega_p)\sigma_{12}^{01}-\gamma_{21}\sigma_{12}^{01}-i\Omega_p\sigma_{32}^{01}+ig_c\sigma_{13}^{00}+i\Omega_s\sigma_{14}^{01}\\

&&\\

\dot{\sigma}_{32}^{01}&=&-i(\omega_{32}-\omega_c)\sigma_{32}^{01}-\gamma_{32}\sigma_{32}^{01}-ig_c(\sigma_{22}^{11}-\sigma_{33}^{00})-i\Omega_{p}\sigma_{12}^{01}+i\Omega_s \sigma_{34}^{01}\\

&&\\

\dot{\sigma}_{14}^{01}&=&i(\omega_{41}+\omega_c-\omega_p-\omega_s)\sigma_{14}^{01}- \gamma_{41} \sigma_{14}^{01} -i\Omega_p\sigma_{34}^{01}+i\Omega_s\sigma_{12}^{01}\\

&&\\

\dot{\sigma}_{34}^{01}&=&i(\omega_{43}+\omega_c-\omega_s)\sigma_{34}^{01}-\gamma_{43} \sigma_{34}^{01} -i\Omega_p \sigma_{14}^{01} - ig_c\sigma_{24}^{11}+i\Omega_s\sigma_{32}^{01} \\

&&\\

\dot{\sigma}_{24}^{11}&=& i(\omega_{42}-\omega_s)\sigma_{24}^{11}-\gamma_{42} \sigma_{24}^{11}-ig_c\sigma_{34}^{01}+i\Omega_s (\sigma_{22}^{11}-\sigma_{44}^{11})\\
&&\\
\end{array}
\label{app:EOM}
\end{equation}
where we introduced the decay rates $\gamma_{31}=\gamma_{e}/2$, $\gamma_{21}=\kappa/2 +\gamma_{v}/2$, $\gamma_{32} = \kappa/2+\gamma_{e}/2$, $\gamma_{43}=\kappa/2+\gamma_{v'}/2$, $\gamma_{42}=\kappa+\gamma_{v'}/2$, and $\gamma_{41}=\gamma_{43}$. The homogeneous probe linewidth is $\gamma_{31}$ and the Raman linewidth is $\gamma_{21}$.

In deriving Eqs. (\ref{app:EOM})-(\ref{app:EOM}), we neglect the contribution of states such as $\ket{2, 0_c}$, $\ket{3, 1_c}$, or  $\ket{4, 0_c}$, which are neither populated nor driven under our imposed assumptions of stationarity and weak signal and probe driving. Accounting for such states would result, for example, in the addition of an extra term proportional to $\sigma_{13}^{11}$ in the right-hand side of equation (\ref{app:EOM}) for $\dot{\sigma}_{13}^{00}$, term that can be shown to vanish in the stationary limit. In other words, the set of Eqs. (\ref{app:EOM}) does not correspond to a complete description of the system coherences, but can be considered as a minimal set of equations of motion that can account for the non-linear optical response of our system of interest. The homogeneous probe susceptibility $\chi_p$ is then obtained by algebraically solving for the steady state probe coherence $\sigma_{13}^{00}(t\rightarrow\infty)$ from the coupled system of equations (\ref{app:EOM}), using the relation $\chi_p=\sigma_{13}^{00}(\infty)/E_p$, which gives Eq. (\ref{eq:chi generic}).

\section{Averaging over Energy Disorder: The Lorentzian Technique}
\label{app:Lorentzian}

Assume that the energies of the states $\ket{1}$,$\ket{ 2}$, $\ket{3}$ and $\ket{4}$ fluctuate as a result of structural disorder (random environment). Then the detunings $\Delta_{31} = \omega_p - \omega_3 + \omega_1$ and $\Delta_{21} = \omega_p - \omega_c - \omega_2 + \omega_1$ also become random quantities. To find the disorder-averaged response, we can numerically integrate Eq. (\ref{eq:chi generic}) in the main text. However, here we will explore an alternative route. Instead of using a realistic Gaussian distribution for molecular levels, we  average the susceptibility over a {\it Lorentzian} distribution of the transition frequencies
\begin{equation}\label{Lorentian}
P_L(x) = \frac{1}{\pi} \frac{\sigma_x}{(x-\langle x \rangle)^2 + \sigma_x^2},
\end{equation}
where $\langle x \rangle$ and $\sigma_x$ are the mean value and the standard deviation of the random variable $x$. The benefit of this approach is that we will get exact analytical averages, which will allow us to gain insight into the system dynamics in simple terms, and help us make meaningful choices of parameters. We show numerically that the results of this procedure compare well with a Gaussian average (see Fig. \ref{fig:disorder scaling}b in the main text).

The proposed technique is based on the observation that there is a class of functions, for which averaging over a Lorentzian distribution can be done instantly. Consider a function of a complex variable $f(z = \Delta+i\gamma)$, where $\Delta$ and $\gamma$ are real quantities; in what follows they will represent, respectively, a random detuning, over which we average, and a constant homogeneous linewidth. We impose two restrictions onto the function $f(z)$: (1) it must decay faster than $z$ for $|z|\to \infty$, and (2) it must not have poles in the upper half-plane. Then averaging of $f(z)$ over a Lorentzian distribution (\ref{Lorentian}) $P_L(\Delta)$ of the detunings with the mean value $\langle \Delta \rangle$ and a standard deviation $\sigma$ writes:

\begin{equation}
\begin{array}{c}
\displaystyle
    \langle f(\Delta+i\gamma) \rangle_{\Delta} = \int\limits_{-\infty}^\infty d\Delta \ f(\Delta+i\gamma) \frac{1}{\pi} \frac{\sigma}{(\Delta - \langle \Delta \rangle )^2 + \sigma^2} = \\

    \\

\displaystyle = \int\limits_{-\infty}^\infty d\Delta \ f(\Delta+i\gamma)\frac{1}{2\pi i}\biggl[ \frac{1}{\Delta-(\langle \Delta \rangle + i \sigma)} - \frac{1}{\Delta-(\langle \Delta \rangle - i \sigma)} \biggr].
\end{array}
\end{equation}

We can calculate this integral by closing the integration contour through the upper half-plane of the complex plane. By assumption (1) the integral over the half-circle vanishes, and by assumption (2) the function $f(z)$ has no poles inside the chosen integration contour, so that the only pole that contributes to the integral is $\Delta = \langle \Delta \rangle + i \sigma$ originating from the Lorentian distribution. The second integral, $\int_{-\infty}^\infty d\Delta~f(\Delta+i\gamma)[\Delta - (\langle \Delta \rangle - i\sigma)]^{-1}/2\pi i$, thus vanishes. By calculating the residue at the only pole $\Delta = \langle \Delta \rangle + i\sigma$ in the first integral and applying the Cauchy theorem, we get:

\begin{equation}\label{trick}
    \langle f(\Delta+i\gamma) \rangle_{\Delta} =  f\biggl(\langle \Delta \rangle + i [\gamma + \sigma]\biggl).
\end{equation}

In summary, as long as the function $f(\Delta +i\gamma)$ satisfies the two criteria listed above, averaging over a Lorentzian distribution consists in replacing the real part of the argument of the function $f(\Delta+i\gamma)$ by its mean value, $\langle \Delta \rangle$, and its imaginary part $\gamma$ by $\gamma+\sigma$, where $\sigma$ is the width of the Lorentian distribution.

Next, we notice that any susceptibility, being considered as a function of complex frequency, must satisfy the criteria (1) and (2), since they are the same requirements that are imposed on all the material functions which satisfy the Kramers-Kronig relations. Hence, averaging the susceptibility over Lorentian distribution of the transition frequencies (or the detunings) with the inhomogeneous broadening $\sigma$ can be done by the replacement (\ref{trick}). The result will be the susceptibility with the fluctuating detuning replaced by its average value, and the homogeneous linewidth $\gamma$ repalced by a sum of the homogeneous and inhomogeneous broadenings, denoted as $\Sigma = \gamma+\sigma$.

Finally, in our multi-level situation, we have to average the susceptibility over more than one Lorentzian distribution. To do this, we repeat this procedure subsequently for each of the disordered transitions. After each averaging we get a new material function that must satisfy the Karmers-Kronig relations, and, consequently, the criteria (1) and (2), so that we can keep repeating this procedure until all the averagings are completed. The result can be symbolicaly written as (here $\Sigma_i = \gamma_i + \sigma_i$):
\begin{equation}\label{averaging}
    \langle \chi(\Delta_1 + i\gamma_1, ..., \Delta_N + i\gamma_N) \rangle_{\Delta_1, ..., \Delta_N} = \chi\biggl(\langle\Delta_1\rangle + i\Sigma_1, ..., \langle\Delta_N\rangle + i\Sigma_N\biggr).
\end{equation}

\end{widetext}

\end{document}